\documentclass[12pt,a4paper]{article}
\usepackage{times}
\usepackage{a4wide}
\usepackage{amsfonts}
\usepackage{amssymb}
\usepackage{amsmath}
\usepackage{ifpdf}
\ifpdf
\usepackage[pdftex,unicode,implicit]{hyperref}
\hypersetup{%
  pdftitle    = {Non-extremal black holes of N=2, d=4 supergravity},
  pdfkeywords = {supergravity, non-extremal black holes},
  pdfauthor   = {Pietro Galli, Tomas Ortin, Jan Perz, Carlos Shahbazi},
  plainpages  = true,
  colorlinks  = true,
  citecolor   = blue,
  urlcolor    = red,
  linkcolor   = black
}
\newcommand{\hepth}[1]{arXiv:{\tt
\href{http://www.arXiv.org/abs/hep-th/#1}{hep-th/#1}}}

\newcommand{\arxiv}[1]{{\tt
\href{http://www.arXiv.org/abs/#1}{arXiv:#1}}}
\else
  \usepackage[dvips]{graphicx}
  \usepackage[unicode,implicit]{hyperref}
  \newcommand{\hepth}[1]{arXiv:{\tt hep-th/#1}}

  \newcommand{\arxiv}[1]{{\tt arXiv:#1}}
\fi
\makeatletter
\@addtoreset{equation}{section}
\makeatother

\begin{document}

\begin{flushright}
\small
IFIC/11-21\\
IFT-UAM/CSIC-11-19\\
\texttt{arXiv:1105.3311 [hep-th]}\\
June 9\textsuperscript{th}, 2011\\
\normalsize
\end{flushright}

\begin{center}

\vspace{1cm}

{\LARGE {\bf Non-extremal black holes\\[.8cm] of $N=2,d=4$ supergravity}}

\vspace{1.5cm}

\begin{center}

\renewcommand{\thefootnote}{\alph{footnote}}
{\sl\large Pietro Galli$^{\dagger}$}
\footnote{E-mail: {\tt Pietro.Galli [at] ific.uv.es}},
{\sl\large Tom\'{a}s Ort\'{\i}n$^{\diamond}$}
\footnote{E-mail: {\tt Tomas.Ortin [at] csic.es}},
{\sl\large Jan Perz$^{\diamond}$}
\footnote{E-mail: {\tt Jan.Perz [at] uam.es}}
{\sl\large and Carlos S.~Shahbazi$^{\diamond}$}
\footnote{E-mail: {\tt Carlos.Shabazi [at] uam.es}}
\renewcommand{\thefootnote}{\arabic{footnote}}

\vspace{.5cm}

${}^{\dagger}${\it Departament de F\'{\i}sica Te\`orica and IFIC (CSIC-UVEG),
Universitat de Val\`encia,\\
C/ Dr.~Moliner, 50, 46100 Burjassot (Val\`encia), Spain}\\

\vspace{.4cm}

${}^{\diamond}${\it Instituto de F\'{\i}sica Te\'orica UAM/CSIC\\
C/ Nicol\'as Cabrera, 13--15,  C.U.~Cantoblanco, 28049 Madrid, Spain}\\

\vspace{.5cm}

\end{center}

{\bf Abstract}

\begin{quotation}

  {\small 
    We propose a generic recipe for deforming extremal black holes into
    non-extremal black holes and we use it to find and study the static
    non-extremal black-hole solutions of several $N=2,d=4$ supergravity models
    ($SL(2,\mathbb{R})/U(1)$, $\overline{\mathbb{CP}}^{n}$ and $STU$ with four
    charges). In all the cases considered, the non-extremal family of solutions
    smoothly interpolates between all the different extremal limits,
    supersymmetric and not supersymmetric. This fact can be used to explicitly
    find extremal non-supersymmetric solutions also in the cases in which the
    attractor mechanism does not completely fix the values of the scalars on
    the event horizon and they still depend on the boundary conditions at
    spatial infinity.

    We compare (supersymmetry) Bogomol'nyi bounds with extremality bounds, we
    find the first-order flow equations for the non-extremal solutions and the
    corresponding superpotential, which gives in the different extremal limits
    different superpotentials for extremal black holes. We also compute the
    \textit{entropies} (areas) of the inner (Cauchy) and outer (event)
    horizons, finding in all cases that their product gives the square of the
    moduli-independent entropy of the extremal solution with the same electric
    and magnetic charges.
}

\end{quotation}

\end{center}

\setcounter{footnote}{0}

\newpage
\pagestyle{plain}

\tableofcontents


\section*{Introduction}

Black holes are among the most interesting objects that occur in theories of
gravity that include or extend general relativity, such as supergravity and
superstring theories because their thermal behavior (Hawking radiation and
Bekenstein--Hawking entropy) provides a unique window into the
quantum-mechanical side of these theories. Their study in the framework of
supergravity and superstring theories has generated a huge body of literature,
the largest part of which concerns extremal (mostly but not always
supersymmetric) black holes.

There are several reasons for having a special interest in extremal black
holes: the solutions are simpler to find, they are protected from classical and
quantum corrections when they are supersymmetric, there is an attractor
mechanism for the scalar fields of most of them
\cite{Ferrara:1995ih,Strominger:1996kf,Ferrara:1996dd,Ferrara:1996um}, their
entropies are easier to interpret microscopically in the framework of
superstring theory \cite{Strominger:1996sh} etc. Much of the progress in their
study has been facilitated by the explicit knowledge of general families of
extremal supersymmetric solutions e.g.~in $N=2,d=4$ supergravity theories,
where we know how to find systematically all of them
\cite{Gibbons:1982fy,Tod:1983pm,Ferrara:1995ih,Behrndt:1996jn,Sabra:1997kq,Sabra:1997dh,Behrndt:1997ny,LopesCardoso:2000qm,Caldarelli:2003pb,Meessen:2006tu,Bellorin:2006xr,Huebscher:2007hj,Cacciatori:2008ek}
(see also the review Ref.~\cite{Mohaupt:2000mj}).

By contrast, only a few non-extremal black-hole solutions are known (for
instance, in $N=2,d=4$ theories), partly because they are more difficult to
find than their extremal counterparts, and partly because they do not enjoy so
many special properties. It is, however, clear that non-extremal black holes
are at least as interesting as the extremal ones from a physical point of view,
because they are closer to those that we may one day be able to
observe. Furthermore, in $d=4$ dimensions adding any amount of angular momentum
to extremal black holes causes the event horizon to disappear
\cite{Bellorin:2006xr}. This does not happen in non-extremal black holes, at
least as long as the angular momentum does not exceed a certain value.

In order to learn more about them it is necessary to have more examples
available for their study. In this paper we are going to propose a procedure to
find non-extremal solutions of $N=2,d=4$ supergravity theories by deforming in
a prescribed way the supersymmetric extremal solutions that we know how to
construct systematically.\footnote{For previous work on near-extremal and
  non-extremal solutions see
  e.g.~Refs.~\cite{Horowitz:1996fn,Cvetic:1996kv,Kastor:1997wi,Behrndt:1997as,Cardoso:2008gm,Gruss:2009kr,Gruss:2009wm}.}
Another prescription has been proposed in the literature, namely the
introduction of an additional harmonic function (called \textit{Schwarzschild
  factor} in Ref.~\cite{Ortin:1996bz} and \textit{non-extremality factor} in
Ref.~\cite{Mohaupt:2010fk}), but it is unclear whether this method will work in
all cases and for all models.

Our proposal makes crucial use of the formalism of Ferrara, Gibbons and Kallosh
in Ref.~\cite{Ferrara:1997tw}, which turns out to be very convenient for our
purposes. This formalism is based on the use of a particular radial coordinate
$\tau$ that covers the exterior of the event horizon (which is always at
$\tau=-\infty$ in these coordinates, a suitable value for the study of
attractors). Furthermore, in this formalism the equations of motion have been
reduced to a very small number of ordinary differential equations in the
variable $\tau$, which should simplify the task of finding solutions.  In these
equations there is a function of the scalars and the electric charges (the
so-called \textit{black-hole potential}), which plays a very important r\^ole,
since its critical points are associated with possible extremal black-hole
solutions. Then, using this formalism, we can also relate more easily the
non-extremal solutions to the extremal solutions that have the same electric
and magnetic charges.

We are going to test our proposal in a number of $N=2,d=4$ models and then
study the main characteristics of the non-extremal solutions constructed. In
this work we consider only regular static black holes.

This paper is organized as follows: in Section~\ref{sec-review} we review
essential facts concerning extremal and non-extremal black holes in the
formalism and coordinates used by Ferrara, Gibbons and Kallosh in
Ref.~\cite{Ferrara:1997tw}. This will help us to establish our notation and
conventions, find an ansatz for the non-extremal black holes based on the
expressions for well-known solutions in these coordinates and show that these
coordinates also cover the region that is bounded by the inner (Cauchy)
horizon. In Section~\ref{sec-axidilaton} we use the ansatz for the
$SL(2,\mathbb{R})/U(1)$ axion-dilaton model to deform the supersymmetric
extremal solutions (which we review first in detail) into non-extremal
solutions, from which we can obtain in adequate limits supersymmetric and
non-supersymmetric extremal black holes. In Section~\ref{sec-CPn} we do the
same for the $\overline{\mathbb{CP}}^{n}$ model. The black hole potential has
flat directions and its non-supersymmetric critical points span a hypersurface
in the moduli space. In other words: the attractor mechanism does not uniquely
fix the values of the scalars on the horizon in terms of the electric and
magnetic charges alone. Consequently the prescription of
Ref.~\cite{Kallosh:2006ib} for constructing full interpolating solutions from
the horizon values of scalars by replacing charges with harmonic functions does
not work. We will find these extremal non-supersymmetric solutions as limits of
the non-extremal ones. In Section~\ref{sec-D0-D4} we do the same for the
well-known 4-charge solutions of the $STU$ model. We show that there are 16
possible extremal limits, and discuss which of them are $N=2$ and/or $N=8$
supersymmetric.  Section~\ref{sec-conclusions} contains our conclusions and
directions for further work.


\section{Extremal and non-extremal black holes}
\label{sec-review}

In this section we are going to review some well-known results on static
extremal and non-extremal black-hole solutions, of which we will make use
later. We will also study some examples of explicit non-extremal solutions in
order to gain insight and formulate a general prescription for the deformation
of supersymmetric extremal solutions into non-extremal solutions.


\subsection{Introductory example: the Schwarzschild black hole}

The prime example of a (non-extremal) black-hole is the Schwarzschild solution,
which in Schwarzschild coordinates is given by

\begin{equation}
ds^{2} =
\left(1-\frac{2M}{r} \right)dt^{2} -\left(1-\frac{2M}{r} \right)^{-1}dr^{2}   
-r^{2} d\Omega_{(2)}^{2}\, ,
\end{equation}

\noindent
where $d\Omega_{(2)}^{2} = d\theta^{2} + \sin^{2}\!\theta\,d\varphi^{2}$ is the
spherically symmetric metric of the unit 2-sphere. In this case, the ``extremal
limit'' is Minkowski spacetime and the non-extremality parameter that goes to
zero in the extremal limit, which we will denote from now on by $r_{0}$, is
just the mass $M$:

\begin{equation}
r_{0}=M\, .
\end{equation}

\noindent
The event horizon is located at the Schwarzschild radius $r_{\rm h}=2M$ and
there is a curvature singularity at $r=0$.

The coordinate transformation

\begin{equation}
r = (\rho +r_{0}/2)^{2}/\rho\, ,  
\end{equation}

\noindent
brings this solution to the spatially isotropic form

\begin{equation}
ds^{2} =   
\left(1-\frac{r_{0}/2}{\rho} \right)^{2}
\left(1+\frac{r_{0}/2}{\rho} \right)^{-2}
dt^{2} -\left(1+\frac{r_{0}/2}{\rho} \right)^{4}
(d\rho^{2}+\rho^{2} d\Omega_{(2)}^{2})\, ,
\end{equation}

\noindent
in which the horizon is located at $\rho_{\rm h}= r_{0}/2$.

In order to study the attractor behavior of different quantities on the event
horizon of a black hole it is convenient to use a radial coordinate $\tau$ such
that $\tau \rightarrow -\infty$ on the horizon.  In the Schwarzschild black
hole there seems to be no attractor behavior, but a coordinate $\tau$ with this
property can be readily found \cite{Gibbons:1982ih}:

\begin{equation}
\rho = -\frac{r_{0}}{2\tanh \frac{r_{0}}{2}\tau}
\end{equation}

\noindent
and with it the Schwarzschild solution can be put in the form

\begin{equation}
\label{eq:generalbhmetric}
\begin{array}{rcl}
ds^{2} 
& = & 
e^{2U} dt^{2} - e^{-2U} \gamma_{\underline{m}\underline{n}}
dx^{\underline{m}}dx^{\underline{n}}\, ,  \\
& & \\
\gamma_{\underline{m}\underline{n}}
dx^{\underline{m}}dx^{\underline{n}}
& = & 
{\displaystyle\frac{r_{0}^{4}}{\sinh^{4} r_{0}\tau}}d\tau^{2} 
+
{\displaystyle\frac{r_{0} ^{2}}{\sinh^{2}r_{0}\tau}}d\Omega^{2}_{(2)}\, ,\\
\end{array}
\end{equation}

\noindent
which is valid for the exterior of any static non-extremal black hole with
different values of the function $U(\tau)$.  For the Schwarzschild black hole

\begin{equation}
U =  r_{0}\tau\, , 
\end{equation}

\noindent
and the radial coordinate $\tau$ takes values in the interval $(-\infty, 0)$,
whose limits correspond to the event horizon and spatial infinity, where the
radius of the 2-spheres becomes infinitely large.  In the interval
$(0,+\infty)$ the above metric describes a Schwarzschild solution with
negative mass and a naked singularity at $\tau\rightarrow +\infty$ (just
transform $\tau\rightarrow -\tau$). In more general cases the interval
$(0,+\infty)$ will describe different patches of the black-hole spacetime.

Using the above general metric for static, non-extremal black holes,
it can be shown \cite{Gibbons:1996af} that the non-extremality
parameter $r_{0}$ satisfies

\begin{equation}
\label{eq:2ST}
r_{0}^{2} = 2ST\, ,  
\end{equation}

\noindent
where $S$ is the Bekenstein entropy and $T$ is the Hawking temperature. 

In the extremal limit $r_{0}\rightarrow 0$ all static black holes are described
by a metric of the same general form of Eq.~(\ref{eq:generalbhmetric}), but the
3-dimensional spatial metric reduces to

\begin{equation}
  \gamma_{\underline{m}\underline{n}}
  dx^{\underline{m}}dx^{\underline{n}}
  =
  \frac{d\tau^{2}}{\tau^{4}}
  +
  \frac{1}{\tau^{2}}d\Omega^{2}_{(2)}\, ,
\end{equation}

\noindent
which, as can be seen by setting $\tau=-1/r$, is the Euclidean metric of
$\mathbb{R}^{3}$ in standard spherical coordinates. In the Schwarzschild case,
$U=0$ in the extremal limit and the full metric becomes Minkowski's.


\subsection{General results}
\label{general}

In Ref.~\cite{Ferrara:1997tw}, in which the attractor behavior of general,
static, $d=4$ black-hole solutions was first studied, it was assumed that all
of them could be written in the general form of
Eq.~(\ref{eq:generalbhmetric}), $U$ being a function of $\tau$ to be
determined and $r_{0}$ (denoted by $c$ in Ref.~\cite{Ferrara:1997tw}) being a
general non-extremality parameter whose value as a function of physical
constants (mass, electric and magnetic charges and asymptotic values of the
scalars) also has to be determined. The action considered in that reference
(slightly adapted to our conventions)\footnote{Our conventions are those of
  Refs.~\cite{Meessen:2006tu,Bellorin:2006xr}.} reads

\begin{equation}
\label{eq:generalaction}
I
=
\int d^{4}x \sqrt{|g|}
\left\{
R +\mathcal{G}_{ij}(\phi)\partial_{\mu}\phi^{i}\partial^{\mu}\phi^{j}
+2 \Im{\rm m} \mathcal{N}_{\Lambda\Sigma}
F^{\Lambda}{}_{\mu\nu}F^{\Sigma\, \mu\nu}
-2 \Re{\rm e} \mathcal{N}_{\Lambda\Sigma}
F^{\Lambda}{}_{\mu\nu}\star F^{\Sigma\, \mu\nu}
\right\}\, ,   
\end{equation}
 
\noindent
and can describe the bosonic sectors of all 4-dimensional ungauged
supergravities for appropriate $\sigma$-model metrics and kinetic matrices
$\mathcal{N}_{\Lambda\Sigma}(\phi)$. The indices $i,j,\dotsc$ run over the
scalar fields and the indices $\Lambda,\Sigma,\dotsc$ over the 1-form
fields. Their numbers are related only for $N\geq 2$ supergravity theories.

Using the general form of the metric for a static non-extremal black
hole, Eq.~(\ref{eq:generalbhmetric}), as well as the conservation of the
electric and magnetic charges, the equations of motion of the above
generic action can be reduced to those of an effective
mechanical system with variables $U(\tau),\phi(\tau)$:

\begin{eqnarray}
\label{eq:e1}
U^{\prime\prime}
+e^{2U}V_{\rm bh}
& = & 0\, ,\\ 
& & \nonumber \\
\label{eq:Vbh-r0-real}
(U^{\prime})^{2} 
+\tfrac{1}{2}\mathcal{G}_{ij}\phi^{i\, \prime}  \phi^{j\, \prime}  
+e^{2U} V_{\rm bh}
& = & r_{0}^{2}\, ,\\
& & \nonumber \\
\label{eq:e3}
(\mathcal{G}_{ij}\phi^{j\, \prime})^{\prime}
-\tfrac{1}{2} \partial_{i}\mathcal{G}_{jk}\phi^{j\, \prime}\phi^{k\, \prime}
+e^{2U}\partial_{i}V_{\rm bh}
& = & 0\, .
\end{eqnarray}

\noindent
Primes signify differentiation with respect to the inverse radial coordinate
$\tau$, which plays the role of the evolution parameter. The so-called
\textit{black-hole potential} is given by\footnote{We adopt the sign of the
  black-hole potential opposite to most of the literature on black-hole
  attractors, conforming instead to the conventions of Lagrangian mechanics.}

\begin{equation}
-V_{\rm bh}(\phi,\mathcal{Q})
\equiv
-\tfrac{1}{2}\mathcal{Q}^{M}\mathcal{Q}^{N} \mathcal{M}_{MN}
\equiv
-\tfrac{1}{2}(p^{\Lambda}\,\,\,\,\, q_{\Lambda})
\left(
  \begin{array}{lr}
    (\mathfrak{I}+\mathfrak{R}\mathfrak{I}^{-1}\mathfrak{R})_{\Lambda\Sigma} &
    -(\mathfrak{R}\mathfrak{I}^{-1})_{\Lambda}{}^{\Sigma} \\
    & \\
    -(\mathfrak{I}^{-1}\mathfrak{R})^{\Lambda}{}_{\Sigma} &
    (\mathfrak{I}^{-1})^{\Lambda\Sigma} \\   
  \end{array}
\right)
\left( 
  \begin{array}{c}
p^{\Sigma} \\
\\
q_{\Sigma} \\
\end{array}
\right)
\, ,
\end{equation}

\noindent
where we replaced each symplectic pair of superscript and subscript indices
$\Lambda,\Sigma,\dotsc$ with a single Latin letter $M,N,\dotsc$, and used the
shorthand

\begin{equation}
\mathfrak{R}_{\Lambda\Sigma} \equiv \Re{\rm e}\mathcal{N}_{\Lambda\Sigma}\, ,
\hspace{1cm}
\mathfrak{I}_{\Lambda\Sigma} \equiv \Im{\rm m}\mathcal{N}_{\Lambda\Sigma}\, ,
\hspace{1cm}
(\mathfrak{I}^{-1})^{\Lambda\Sigma}\mathfrak{I}_{\Sigma\Gamma} = \delta^{\Lambda}{}_{\Gamma}\, .
\end{equation}

Eqs.~(\ref{eq:e1}) and (\ref{eq:e3}), but not the constraint
Eq.~(\ref{eq:Vbh-r0-real}), can be derived from the effective
action\footnote{The three equations \eqref{eq:e1}--\eqref{eq:e3} can be derived
  from a more general effective action, which is reparametrization invariant:
\begin{equation}
  I_{\rm eff}[U,\phi^{i},e] = \int d\tau \left\{ 
    e^{-1}\left[ (U^{\prime})^{2}  
      +\tfrac{1}{2}\mathcal{G}_{ij}\phi^{i\, \prime}  \phi^{j\, \prime}  \right]
    -e\left[e^{2U} V_{\rm bh} -r_{0}^{2}\right]
  \right\}\, ,  
  \end{equation}
  where $e(\tau)$ is an auxiliary einbein. We can recover the effective action
  Eq.~(\ref{eq:effectiveaction}) in the gauge $e(\tau)=1$, in which the
  equation of motion of $e$ gives precisely the constraint
  Eq.~(\ref{eq:Vbh-r0-real}). The constant term in
  Eq.~\eqref{eq:effectiveaction} is usually ignored, as it is a total
  derivative.
}

\begin{equation}
\label{eq:effectiveaction}
  I_{\rm eff}[U,\phi^{i}] = \int d\tau \left\{ 
(U^{\prime})^{2}  
+\tfrac{1}{2}\mathcal{G}_{ij}\phi^{i\, \prime}  \phi^{j\, \prime}  
-e^{2U} V_{\rm bh} +r_{0}^{2}
  \right\}\, .  
\end{equation}

In Ref.~\cite{Ferrara:1997tw} it was shown that for regular extremal
($r_{0}=0$) black holes the values of the scalars on the event horizon
$\phi^{i}_{\rm h}$ are critical points of the black hole potential,\footnote{In
  the absence of stationary points the scalars would be singular on the
  horizon. We do not consider such cases.}  i.e.~they satisfy

\begin{equation}
\label{eq:criticalvalues}
  \left. \partial_{i}V_{\rm bh}   \right\rvert_{\phi_{\rm h}}=0\, .
\end{equation}

\noindent
These equations can be solved in terms of the charges but, if the black hole
potential has \textit{flat directions}, the equations will be underdetermined
and their solution will have residual dependence on the asymptotic values of
the scalars at spatial infinity ($\tau\rightarrow 0^{-}$):

\begin{equation}
\phi_{\rm h}=\phi_{\rm h}(\phi_{\infty},\mathcal{Q})\, .
\end{equation}

\noindent
Furthermore, it was shown that the value of the black-hole potential at the
critical points gives the entropy:

\begin{equation}
S = -\pi V_{\rm bh}(\phi,\mathcal{Q})\bigr\rvert_{\phi_{\rm h}}
\end{equation}

\noindent
and that the near-horizon geometry is that of $AdS_{2}\times S^{2}$ with the
$AdS_{2}$ and $S^{2}$ radii both equal to $(-V_{\rm bh}\rvert_{\phi_{\rm
    h}})^{1/2}$. Even though the critical loci may not be isolated points, in
which case the scalars will vary along the flat directions of the potential when
one changes $\phi_{\infty}$, the stationary value itself will not be affected,
hence the entropy depends on the charges only \cite{Sen:2005iz}. Each solution
to Eq.~(\ref{eq:criticalvalues}) yields a possible set of values of the scalars
on the event horizon and of the radii, thus a possible extremal black-hole
solution.

In the general case one can prove the following extremality bound
\cite{Ferrara:1997tw}:

\begin{equation}
\label{eq:generalbound}
r_{0}^{2} = M^{2}+ \tfrac{1}{2}\mathcal{G}_{ij}(\phi_{\infty})
\Sigma^{i}\Sigma^{j}   +V_{\rm bh}(\phi_{\infty},\mathcal{Q})
\geq 0\, ,
\end{equation}

\noindent
where $M,\Sigma^{i}$ are the mass and scalar charges defined by the behavior
at spatial infinity ($\tau\rightarrow 0^{-}$)

\begin{equation}
  \begin{array}{rcl}
U & \sim & 1 +M\tau\, ,\\
& & \\
\phi^{i} & \sim & \phi^{i}_{\infty} -\Sigma^{i}\tau\, .\\
\end{array}
\end{equation}


\subsubsection{Flow equations}

Whenever the potential term can be represented as a sum of squares of
derivatives of a so-called \textit{(generalized) superpotential} function
$Y(U,\phi^{i},\mathcal{Q},r_{0})$ of the warp factor $U$ and the scalars
$\phi^{i}$,

\begin{equation}
-\left[e^{2U}V_{\rm bh}-r_{0}^{2} \right] = 
(\partial_{U}Y)^{2} +2\,\mathcal{G}^{ij}\partial_{i}Y\partial_{j}Y\, ,
\end{equation}

\noindent
the effective action Eq.~(\ref{eq:effectiveaction}) also admits a rewriting as a
sum of squares (up to a total derivative)

\begin{equation}
\label{eq:effectiveaction2}
  I_{\rm eff}[U,\phi^{i}] = \int d\tau \left\{ 
(U^{\prime} \pm \partial_{U}Y)^{2}  
+\tfrac{1}{2}\mathcal{G}_{ij}
(\phi^{i\, \prime} \pm 2\,\mathcal{G}^{ik}\partial_{k}Y)
(\phi^{j\, \prime} \pm 2\,\mathcal{G}^{jl}\partial_{l}Y)
\mp 2 Y^{\prime}
  \right\}\, ,
\end{equation}

\noindent
whose variation leads to first-order gradient flow equations, solving the
second-order equations of motion \cite{Perz:2008kh}:\footnote{This generalized
  the results of Refs.~\cite{Miller:2006ay,Janssen:2007rc}. For first-order
  equations with a $\tau$-dependent superpotential see
  Refs.~\cite{Andrianopoli:2007gt,Andrianopoli:2009je}.}

\begin{eqnarray}
\label{eq:U-Y}
U^{\prime} & = & \partial_{U}Y\, ,\\
& & \nonumber \\
\label{eq:phi-Y}
\phi^{i\, \prime} & = & 2\,\mathcal{G}^{ij}\partial_{j}Y\, .
\end{eqnarray}

\noindent
Of the two signs in Eq.~(\ref{eq:effectiveaction2}), only one,
dependent on conventions, is physically admissible. We take $\partial_{U} Y$
to be positive.) It is easy to see that

\begin{equation}
  \partial_{i}Y = 0\,\,\, \Rightarrow\,\,\, \partial_{i}V_{\rm bh} =0\, ,
\end{equation}

\noindent
which sometimes simplifies the task of finding critical points of the
black-hole potential. Observe also that when there is a generalized
superpotential $Y$, the mass and scalar charges are determined by its
derivatives at spatial infinity $\tau \rightarrow 0^{-}$:

\begin{equation}
\label{eq:MSigma}
M =  \lim_{\tau \rightarrow 0^{-}} \partial_{U}Y\, ,
\hspace{1cm}
\Sigma^{i} = - \lim_{\tau \rightarrow 0^{-}} \mathcal{G}^{ij}\partial_{j}Y\, .
\end{equation}

The generalized superpotential $Y(U,\phi^{i},\mathcal{Q},r_{0})$ has been proven
\cite{Andrianopoli:2009je,Chemissany:2010zp} to exist in theories whose scalar
manifold (after timelike dimensional reduction) is a symmetric coset space, thus
in particular for extended supergravities with more than $8$ supercharges.

In the extremal cases, when there is a generalized superpotential function
$Y(U,\phi,\mathcal{Q})$, it factorizes into

\begin{equation}
Y(U,\phi,\mathcal{Q}) = e^{U}W(\phi,\mathcal{Q})\, ,
\end{equation}

\noindent
where $W$ is called the \textit{superpotential}. The flow equations take the
form \cite{Ceresole:2007wx}

\begin{eqnarray}
\label{eq:U-W}
U^{\prime} & = & e^{U}W\, ,\\
& & \nonumber \\
\label{eq:phi-W}
\phi^{i\, \prime} & = & 2\, e^{U} \mathcal{G}^{ij}\partial_{j}W\, .
\end{eqnarray}

\noindent
In supergravities with more than $8$ supercharges and in the extremal limit
there is always at least one superpotential associated with the skew
eigenvalues of the central charge, the above flow equations are related to the
Killing spinor identities, and the corresponding extremal black-hole solutions
are supersymmetric. However, in general there are extremal black-hole solutions
that are not supersymmetric and satisfy the above flow equations for a
different superpotential. We will discuss this point in more detail for $N=2$
supergravity in the next section.

The stationary values of the superpotential
\begin{equation}
\left .\partial_{i}W \right|_{\phi_{\rm h}}=0\,
\end{equation}

\noindent
give the the entropy:

\begin{equation}
\label{eq:Wentropy}
S = \pi |W(\phi_{\rm h},\mathcal{Q})|^{2}\, .  
\end{equation}


\subsection{$N=2,d=4$ supergravity}
\label{susysolutions}

In this paper we will focus on theories of ungauged $N=2,d=4$ supergravity
coupled to $n$ vector supermultiplets (that is, with $\bar{n}=n+1$ vector
fields $A^{\Lambda}{}_{\mu}$, $\Lambda=0,1,\dotsc,n$, taking into account the
graviphoton).\footnote{See, for instance, Ref.~\cite{Andrianopoli:1996cm}, the
  review \cite{kn:toinereview}, and the original works
  \cite{deWit:1984pk,deWit:1984px} for more information on $N=2,d=4$
  supergravities.} The $n$ scalars of these theories, denoted by $Z^{i}$,
$i=1,\dotsc,n$ are complex and parametrize a special K\"ahler manifold with
K\"ahler metric $\mathcal{G}_{ij^{*}}=\partial_{i}\partial_{j^{*}}\mathcal{K}$,
where $\mathcal{K}(Z,Z^{*})$ is the K\"ahler potential, and the
Eqs.~(\ref{eq:e1})--(\ref{eq:e3}) can be rewritten in the form

\begin{eqnarray}
\label{eq:1}
U^{\prime\prime}
+e^{2U}V_{\rm bh}
& = & 0\, ,\\ 
& & \nonumber \\
\label{eq:2}
(U^{\prime})^{2} 
+\mathcal{G}_{ij^{*}}Z^{i\, \prime}  Z^{*\, j^{*}\, \prime}  
+e^{2U} V_{\rm bh}
& = & r_{0}^{2}\, ,\\
& & \nonumber \\
\label{eq:3}
Z^{i\, \prime\prime}
+\mathcal{G}^{ij^{*}}\partial_{k}\mathcal{G}_{lj^{*}}Z^{k\, \prime}Z^{l\, \prime}
+e^{2U}\mathcal{G}^{ij^{*}}\partial_{j^{*}}V_{\rm bh}
& = & 0\, .
\end{eqnarray}

\noindent
Furthermore, the black-hole potential takes the simple form

\begin{equation}
-V_{\rm bh}(Z,Z^{*},\mathcal{Q}) = |\mathcal{Z}|^{2} 
+\mathcal{G}^{ij^{*}}\mathcal{D}_{i}\mathcal{Z}\mathcal{D}_{j^{*}}\mathcal{Z}^{*}\, ,
\end{equation}
 
\noindent
where 

\begin{equation}
\mathcal{Z}=\mathcal{Z}(Z,Z^{*},\mathcal{Q})  \equiv 
\langle \mathcal{V}\mid\mathcal{Q} \rangle =
-\mathcal{V}^{M}\mathcal{Q}^{N}\Omega_{MN}
=
p^{\Lambda}\mathcal{M}_{\Lambda} -q_{\Lambda}\mathcal{L}^{\Lambda}\, ,
\end{equation}

\noindent
is the \textit{central charge} of the theory,
$\mathcal{V}^{M}=(\mathcal{L}^{\Lambda},\mathcal{M}_{\Lambda})$ is the
covariantly holomorphic symplectic section, $(\Omega_{MN}) = \left(
  \begin{array}{cc}
   0 & \mathbb{I}_{\bar{n}\times \bar{n}} \\ - \mathbb{I}_{\bar{n}\times \bar{n}} & 0 \\ 
  \end{array}
\right)$ is the symplectic metric, and 

\begin{equation}
\mathcal{D}_{i}\mathcal{Z} = 
e^{-\mathcal{K}/2} \partial_{i}\left(e^{\mathcal{K}/2}\mathcal{Z} \right)\, ,  
\end{equation}

\noindent
is the K\"ahler covariant derivative.

Since $\mathcal{D}_{i}\mathcal{Z} = 2
(\mathcal{Z}/\mathcal{Z}^{*})^{1/2} \partial_{i}|\mathcal{Z}|$, in
$N=2$ theories there is always at least one superpotential

\begin{equation}
W =|\mathcal{Z}|\, ,  
\end{equation}

\noindent
and the associated flow equations (\ref{eq:U-W}), (\ref{eq:phi-W}) for extremal
black holes take the form

\begin{eqnarray}
\label{eq:U-|Z|}
U^{\prime} & = & e^{U}|\mathcal{Z}|\, ,\\
& & \nonumber \\
\label{eq:Z-|Z|}
Z^{i\, \prime} & = & 2e^{U}\mathcal{G}^{ij^{*}}\partial_{j^{*}}|\mathcal{Z}|\, .
\end{eqnarray}

It can be shown that these flow equations follow from the $N=2$ Killing spinor
identities and the corresponding extremal black-hole solutions are
supersymmetric.\footnote{For a rigorous proof, see
  Ref.~\cite{Bellorin:2006xr}.} $|\mathcal{Z}|$ is the only superpotential
associated to supersymmetric solutions in $N=2$ theories, but there can be
more non-supersymmetric superpotentials $W$.

Then, for $N=2$ theories, the critical points of the black-hole potential
(that we will loosely call \textit{attractors} from now on) are of two kinds:

\begin{description}
\item[Supersymmetric (or \textit{BPS}) attractors]\hspace{-1ex}, for which

\begin{equation}
\mathcal{D}_{i}\mathcal{Z} \bigr|_{Z_{\rm h}}=0
\qquad\text{or, equivalently}\qquad
\partial_{i}|\mathcal{Z}| \Bigr|_{Z_{\rm h}}=0\, .
\end{equation}

\noindent
As we have mentioned, the extremal black-hole solutions associated to these
attractors are supersymmetric and the functions $U(\tau),Z^{i}(\tau)$ satisfy
the above flow equations. Furthermore, according to the general results, the
entropy is given by the value of the central charge at the horizon

\begin{equation}
S = \pi|\mathcal{Z}(Z_{\rm h},Z^{*}_{\rm h},\mathcal{Q})|^{2}  
\end{equation}

\noindent
and the mass of the black hole is given by the value of the central charge at
infinity (BPS relation)

\begin{equation}
M= |\mathcal{Z}(Z_{\infty},Z^{*}_{\infty},\mathcal{Q})|\, .  
\end{equation}

In this case, since at supersymmetric critical points the Hessian of the black
hole potential $-V_{\rm bh}$ is proportional to the (positive definite) metric
on the scalar manifold, these points must be minima \cite{Ferrara:1997tw}. As a
consequence, the scalars on the horizon take \textit{attractor values} $Z_{\rm
  h}=Z_{\rm h}(\mathcal{Q})$, determined only by the electric and magnetic
charges and independent of the asymptotic boundary conditions (at least within
a single ``basin of attraction'' \cite{Moore:1998pn}). To put it differently:
supersymmetric attractors are stable. As already remarked, the attractor
mechanism may fail for certain choices of charges for which the horizon is
singular (\textit{small black holes}).

\item[Non-supersymmetric attractors]
  \cite{Sen:2005wa,Goldstein:2005hq,Tripathy:2005qp}.
They satisfy an equation of the form

\begin{equation}
\left. \partial_{i}W \right|_{Z_{\rm h}}=0\, ,  
\end{equation}

\noindent
for a superpotential function $W(Z,Z^{*},\mathcal{Q}) \neq |\mathcal{Z}|$
\cite{Ceresole:2007wx}, and the solution satisfies the corresponding flow
equations (\ref{eq:U-W}), (\ref{eq:phi-W}). The entropy will be given by
Eq.~(\ref{eq:Wentropy}) and the mass and scalar charges by
Eqs.~(\ref{eq:MSigma}):

\begin{equation}
S = \pi |W(Z_{\rm h},Z^{*}_{\rm h},\mathcal{Q})|^{2}\, ,  
\hspace{.5cm}
M = |W(Z_{\infty},Z^{*}_{\infty},\mathcal{Q})|\, ,
\hspace{.5cm}
\Sigma^{i} =  
-\mathcal{G}^{ij}\partial_{j}W(Z_{\infty},Z^{*}_{\infty},\mathcal{Q})\, .
\end{equation}

One of the main differences with the supersymmetric case is that the stationary
points of the black hole potential do not necessarily need to be minima. For
models whose scalar manifold is a homogeneous space (in particular thus for all
models embeddable in $N > 2$ supergravity) the Hessian at these points
(expressed in a real basis \cite{Bellucci:2006ew, Cardoso:2006cb}), has
non-negative eigenvalues, therefore such stationary points are also stable, but
only up to possible flat directions \cite{Ferrara:2007pc, Ferrara:2007tu}. It
means that the attractor mechanism is no longer guaranteed to completely fix the
values of the scalars on the horizon $Z^{i}_{\rm h}$, which may still depend on
the asymptotic values $Z^{i}_{\infty}$ as well as on the charges $\mathcal{Q}$,
even though the entropy will only depend on the charges. In this sense one may
speak of \textit{moduli spaces of attractors} parametrized by (combinations of)
the $Z^{i}_{\infty}$, as opposed to the supersymmetric attractors, which are
isolated points in the target space of the scalars.

\end{description}

Only in the supersymmetric case $\Sigma^{i}= \left.
  \mathcal{D}^{i}\mathcal{Z}^{*} \right|_{Z_{\infty}}$ and, therefore, the
general extremality bound Eq.~(\ref{eq:generalbound}) does not reduce to just
the BPS bound $r_{0}^{2}= M^{2}
-|\mathcal{Z}(Z_{\infty},Z^{*}_{\infty},\mathcal{Q})|^{2}\geq 0$ (otherwise, all
extremal black holes in $N=2$ supergravity would automatically be
supersymmetric, which is not true). One of our goals is to study the general
extremality bound and interpret it in terms of the central charge and other
known quantities, explaining why and how it happens that supersymmetry always
implies extremality, but not the other way around, as first shown in
Ref.~\cite{Khuri:1995xq} (see also \cite{Ortin:1997yn}).


\subsubsection{$N=2,d=4$ black-hole solutions}

How are the complete black-hole solutions (or, equivalently, the variables
$U(\tau),Z^{i}(\tau)$) found? For supersymmetric (and, therefore, extremal)
$N=2$ supergravity solutions there is a well-established method to construct
systematically all the possible black-hole solutions
\cite{Ferrara:1995ih,Strominger:1996kf,Sabra:1997kq,Sabra:1997dh,Behrndt:1997ny,LopesCardoso:2000qm,Denef:2000nb,Bates:2003vx,Meessen:2006tu}. We
will follow the prescription given in Ref.~\cite{Meessen:2006tu}:

\begin{enumerate}
\item Introduce a complex function $X(Z,Z^{*})$ with the same K\"ahler weight as
  the canonical symplectic section $\mathcal{V}$ so that the quotient
  $\mathcal{V}/X$ is invariant under K\"ahler transformations.\footnote{This
    prescription does not depend on the K\"ahler gauge. A function playing the
    same role as $X$, namely $1/\mathcal{Z}^{*}$, was also introduced in
    Ref.~\cite{Behrndt:1996jn}.}

\item Define the real symplectic vectors $\mathcal{R}$ and $\mathcal{I}$ by

\begin{equation}
\mathcal{R}+i\mathcal{I}\equiv \mathcal{V}/X\, .  
\end{equation}

The components of $\mathcal{R}$ can always be expressed in terms of those of
$\mathcal{I}$ (by solving the \textit{stabilization equations} of
Refs.~\cite{Sabra:1997dh,Behrndt:1997ny}),\footnote{Since the relations must
  remain the same at all points in space, it suffices to infer them on the
  horizon, where the stabilization equations reduce to the \textit{attractor
    equations} of Ref.~\cite{Ferrara:1996dd}.} although in some cases the
relations may be difficult to find explicitly.

\item The $2\bar{n}$ components of imaginary part $\mathcal{I}^{M}$ are given by
  as many real harmonic functions in $\mathbb{R}^{3}$.  For single-center,
  spherically symmetric, black-hole solutions, they must have the
  form\footnote{The factor $1/\sqrt{2}$ is required for the correct
    normalization of the charges (in particular, to have the same normalization
    of the charges used in the definition of the black-hole potential) and it
    was omitted in Ref.~\cite{Bellorin:2006xr}.}
      
\begin{equation}
\label{eq:generalharmonicfunction}
\mathcal{I}^{M} = \mathcal{I}^{M}_{\infty}
-\tfrac{1}{\sqrt{2}}\mathcal{Q}^{M}\tau\, .
\end{equation}

Furthermore, in order not to have NUT charge (and have staticity) we must
require \cite{Bellorin:2006xr}

\begin{equation}
\label{eq:noNUT}
\langle \mathcal{I}_{\infty} \mid\mathcal{Q} \rangle =
-\mathcal{I}^{M}_{\infty}\mathcal{Q}^{N}\Omega_{MN}=0\, .  
\end{equation}

The choice of $\mathcal{I}^{M}$ determines the components $\mathcal{R}^{M}$
according to the pervious discussion.

\item The scalar fields are given by 

\begin{equation}
Z^{i} 
=\frac{\mathcal{L}^{i}}{\mathcal{L}^{0}} 
=\frac{\mathcal{L}^{i}/X}{\mathcal{L}^{0}/X} 
=  \frac{\mathcal{R}^{i}+i\mathcal{I}^{i}}{\mathcal{R}^{0}+i\mathcal{I}^{0}}\, ,
\end{equation}

\noindent
and the metric function $U$ is given by 

\begin{equation}
\label{eq:metricfunction}
e^{-2U} = 
\frac{1}{2|X|^{2}}
=\langle\, \mathcal{R} \mid \mathcal{I}\, \rangle =
-\mathcal{R}^{M}\mathcal{I}^{N}\Omega_{MN}\, .
\end{equation}

We will not need the explicit form of the vector fields but they can be found
in Ref.~\cite{Bellorin:2006xr}.

\end{enumerate}

Some extremal but non-supersymmetric solutions can be constructed from the
attractor values \cite{Kallosh:2006bt,Kallosh:2006ib} by replacing the electric
and magnetic charges in the expressions for the scalars on the horizon by
harmonic functions (the metric function is obtained in the same way from the
entropy). It is not clear, however, that this is always applicable, in
particular when there are moduli spaces of non-supersymmetric attractors, as in
the $\overline{\mathbb{CP}}^{n}$ model (Section \ref{sec-CPn}).

There is no general algorithm to construct non-extremal black-hole solutions
either. In some cases, the introduction of an additional harmonic function
(called \textit{Schwarzschild factor} in Ref.~\cite{Ortin:1996bz} and
\textit{non-extremality factor} in Ref.~\cite{Mohaupt:2010fk}) appears to be
enough, but the explicit non-extremal solution \cite{LozanoTellechea:1999my}
seems to suggest that this prescription may not always work. In order to gain
more insight into this problem, which is of our main interest in this paper, we
are going to examine in detail more examples of non-extremal solutions. Then we
will formulate a prescription to deform any static extremal supersymmetric
black-hole solution of $N=2,d=4$ supergravity into a non-extremal one and, next,
we will apply it to several examples in the following sections.


\subsection{Second example: the Reissner--Nordstr\"om black hole}
\label{RNbh}

Let us consider pure $N=2,d=4$ supergravity, with the bosonic action

\begin{equation}
I = \int d^{4}x \sqrt{|g|}
\left[
R -F^{2} \right]\, ,
\end{equation}

\noindent
which corresponds to a canonical section and period matrix 

\begin{equation}
\mathcal{V}= 
\left( 
\begin{array}{c}
\mathcal{L}^{0}\\
\mathcal{M}_{0} \\    
\end{array}
\right)  
= 
\left( 
\begin{array}{c}
i\\
\frac{1}{2} \\    
\end{array}
\right)\, ,
\hspace{1cm}  
\mathcal{N}_{00} = -\frac{i}{2}\, .
\end{equation}

\noindent
The central charge and black-hole potential are

\begin{equation}
\mathcal{Z} = \tfrac{1}{2}p -iq\, ,
\hspace{1cm}
-V_{\rm bh} = |\mathcal{Z}|^{2}\, ,
\end{equation}

\noindent
and, since there are no scalars, it has no critical points. 

The supersymmetric extremal black-hole solutions can be constructed using the
mentioned algorithm of Ref.~\cite{Meessen:2006tu}. First, we introduce the
function $X$ and the two harmonic functions

\begin{equation}
  \begin{array}{rcl}
\mathcal{I}^{0} & = & \Im {\rm m}(\mathcal{L}^{0}/X) = \mathcal{I}^{0}_{\infty} 
-{\displaystyle\frac{p^{0}}{\sqrt{2}}}\tau\, ,\\
& & \\  
\mathcal{I}_{0} & = & \Im {\rm m}(\mathcal{M}_{0}/X) = \mathcal{I}_{0\, \infty} 
-{\displaystyle\frac{q_{0}}{\sqrt{2}}}\tau\, ,\\
\end{array}
\end{equation}

\noindent
where $\mathcal{I}^{0}_{\infty}, \mathcal{I}_{0\, \infty}$ are constants to be
determined later.\footnote{These constants are often set equal to $1$ from the
  beginning, which is in general incorrect, as we are going to show.}  It is
convenient to combine these two real harmonic functions into a single complex
harmonic function

\begin{equation}
  \mathcal{H} \equiv \tfrac{1}{\sqrt{2}} (\mathcal{I}^{0}+2i\mathcal{I}_{0})
  = \mathcal{H}_{\infty} - \mathcal{Z}\tau\, .
\end{equation}

\noindent
Then, it is easy to see that the zero-NUT-charge condition
Eq.~(\ref{eq:noNUT}) can be written in the form

\begin{equation}
\label{eq:nonNUT-RN}
N=\Im {\rm m}(\mathcal{H}_{\infty}\mathcal{Z}^{*}) =0\, .  
\end{equation}

The stabilization equations determine the real parts

\begin{equation}
  \begin{array}{rcl}
\mathcal{R}^{0} & = & -2\mathcal{I}_{0}\, ,\\
& & \\  
\mathcal{R}_{0} & = & \frac{1}{2}\mathcal{I}^{0}\, ,\\
\end{array}
\end{equation}

\noindent
and then the metric function is given by 

\begin{equation}
e^{-2U} = |\mathcal{H}|^{2} = |\mathcal{H}_{\infty}|^{2} -2\Re{\rm
  e}(\mathcal{H}_{\infty}\mathcal{Z}^{*})\tau +|\mathcal{Z}|^{2}\tau^{2}\, .  
\end{equation}

\noindent
Asymptotic flatness requires $|\mathcal{H}_{\infty}|^{2}=1$ and indicates that
$M=\Re{\rm e}(\mathcal{H}_{\infty}\mathcal{Z}^{*})$, and then we get the
well-known extremal, dyonic, Reissner--Nordstr\"om (RN) solution:

\begin{equation}
\mathcal{H}_{\infty} = \frac{\mathcal{Z}}{|\mathcal{Z}|}\, ,
\hspace{1cm}
M=|\mathcal{Z}|\, , 
\hspace{1cm}
S = \pi |\mathcal{Z}|^{2}\, .
\hspace{1cm}
e^{-2U} = (1-|\mathcal{Z}|\tau)^{2}\, .
\end{equation}

\noindent 
Observe that $e^{-2U}$ ends up as the square of a real harmonic function, which
we can call $H$.

The non-extremal RN solutions are, of course, known as well. In our
conventions, and Schwarzschild-like coordinates, the metric takes the form

\begin{equation}
ds^{2} = \frac{(r-r_{+})(r-r_{-})}{r^{2}}dt^{2}  
-\frac{r^{2}}{(r-r_{+})(r-r_{-})}dr^{2} -r^{2}d\Omega^{2}_{(2)}\, ,
\end{equation}

\noindent
where 

\begin{equation}
r_{\pm} = M \pm r_{0}\, ,  
\end{equation}

\noindent
are the values of $r$ at which the outer (event) horizon ($+$) and inner
(Cauchy) horizon ($-$) are located, and 

\begin{equation}
r^{2}_{0} = M^{2}-|\mathcal{Z}|^{2}\, ,
\end{equation}

\noindent
is the non-extremality parameter.

In order to study this solution using the black-hole potential formalism we
first need to reexpress it in terms of the coordinate $\tau$. As an
intermediate step we reexpress it in terms of spatially isotropic coordinates

\begin{equation}
r = [\rho^{2} +M\rho +r^{2}_{0}/4]/\rho\, ,
\end{equation}

\noindent
so it takes the form ($\rho_{\pm} \equiv M\pm |\mathcal{Z}|$)

\begin{equation}
ds^{2} 
=
\frac{\left(1-\frac{r_{0}/2}{\rho} \right)^{2}\left(1+\frac{r_{0}/2}{\rho}
\right)^{2}}{\left(1+\frac{\rho_{+}/2}{\rho} \right)^{2}
\left(1+\frac{\rho_{-}/2}{\rho} \right)^{2}}
dt^{2} 
-
\left(1+\frac{\rho_{+}/2}{\rho} \right)^{2}\left(1+\frac{\rho_{-}/2}{\rho} \right)^{2}
(d\rho^{2}+\rho^{2} d\Omega_{(2)}^{2})\, .
\end{equation}

\noindent
For $M=|\mathcal{Z}|$ ($\rho_{-}=r_{0}=0$) we recover the extremal solution
just studied (with $\rho=-1/\tau$). Next, we change to the coordinate $\tau$ as
in the Schwarzschild case with $M$ replaced by $r_{0}$

\begin{equation}
\rho = -\frac{r_{0}}{2\tanh \frac{r_{0}}{2}\tau}\, ,
\end{equation}

\noindent
to obtain a metric of the standard form Eq.~(\ref{eq:generalbhmetric}) with 

\begin{equation}
e^{-2U} = e^{-2r_{0}\tau} 
\left[\frac{r_{+}}{2r_{0}} -\frac{r_{-}}{2r_{0}}e^{2r_{0}\tau}  
\right]^{2}\, .
\end{equation}

This metric function contains a \textit{Schwarzschild factor}
$e^{-2r_{0}\tau}$, which is the only one that remains when the charge vanishes,
and the square of a function which is not a harmonic function in
$\mathbb{R}^{3}$ but can be seen as a deformation of the function $H=
1-|\mathcal{Z}|\tau$:

\begin{equation}
\lim_{r_{0}\rightarrow 0}
\left[\frac{r_{+}}{2r_{0}} -\frac{r_{-}}{2r_{0}}e^{2r_{0}\tau} \right] 
= H\, .  
\end{equation}

\noindent
As in the Schwarzschild case, when the radial coordinate coordinate $\tau$
takes values in the interval $(-\infty, 0)$, whose limits correspond to the
event horizon and spatial infinity, the metric covers the exterior of the
horizon. The explicit relation between the original Schwarzschild-like radial
coordinate $r$ and $\tau$ in that interval is

\begin{equation}
\tau = \frac{2}{r_{0}} \mathrm{arctanh}\,  
\left\{ 
\frac{-r_{0}}{(r-M) +\sqrt{(r-M)^{2} -r_{0}^{2}}} 
\right\}\, ,
\hspace{1cm}
r \in (r_{+},+\infty)\, .
\end{equation}

In the RN case, however, \textit{the same metric} also covers the interior of
the inner horizon when $\tau$ takes values in the interval $(\frac{2}{r_{0}}
\mathrm{arctanh}\, \sqrt{\frac{M-|\mathcal{Z}|}{M+|\mathcal{Z}|}}, +\infty)$,
whose limits correspond to the singularity at the origin and the inner
horizon. The explicit relation between the original Schwarzschild-like radial
coordinate $r$ and $\tau$ in that interval is

\begin{equation}
\tau = \frac{2}{r_{0}} \mathrm{arctanh}\,  
\left\{ 
\frac{-r_{0}}{(r-M) -\sqrt{(r-M)^{2} -r_{0}^{2}}} 
\right\}\, ,
\hspace{1cm}
r \in (0,r_{-})\, .
\end{equation}

It is easy to see that $e^{2U}$ tends to zero in the two limits
$\tau\rightarrow \pm \infty$ and that the coefficient of $d\Omega^{2}_{(2)}$
in the metric, which can be understood as the square radius of the spatial
sections of the horizons 

\begin{equation}
\frac{(2r_{0})^{2}e^{-2U}}{(e^{r_{0}\tau} -e^{-r_{0}\tau})^{2}}=
\left(
{\displaystyle\frac{r_{\pm}-r_{\mp}e^{\pm 2r_{0}\tau}}{e^{\pm 2r_{0}\tau} -1}}
\right)^{2} 
\stackrel{\tau\rightarrow \mp \infty}{\longrightarrow}  r_{\pm}^{2}\, .
\end{equation}

This allows us to compute the areas and, therefore, the ``entropies''
associated with both horizons using the standard metric:

\begin{equation}
  S_{\pm}/\pi = (r_{\pm})^{2}\, ,  
\end{equation}

\noindent
and, using the general result Eq.~(\ref{eq:2ST}), the temperatures

\begin{equation}
T_{\pm} = \frac{r_{0}^{2}}{2S_{\pm}}= \frac{1}{2 \pi}(r_{0}/r_{\pm})^{2}\, .
\end{equation}


\subsection{General prescription}\label{generalprescription}

The previous result suggests the following prescription for deforming
extremal, static, supersymmetric solutions of $N=2,d=4$ supergravity into
non-extremal solutions: if the supersymmetric solution is given by

\begin{equation}
U(\tau) = U_{\rm e}[H(\tau)]\, ,
\hspace{1cm}
Z^{i}(\tau) = Z^{i}_{\rm e}[H(\tau)]\, ,
\end{equation}

\noindent
where $U_{\rm e}$ and $Z^{i}_{\rm e}$ are the functions of certain harmonic
functions $H_{\alpha}(\tau)= H_{\alpha\, \infty}-Q_{\alpha}\tau$ ($\alpha$
being some index) that one finds following the standard prescription for
supersymmetric black holes, then the non-extremal solution is given by

\begin{equation}
U(\tau) = U_{\rm e}[\hat{H}(\tau)]+r_{0}\tau\, ,
\hspace{1cm}
Z^{i}(\tau) = Z^{i}_{\rm e}[\hat{H}(\tau)]\, ,  
\end{equation}

\noindent
where the harmonic functions $H$ have been replaced by the hatted functions

\begin{equation}
\hat{H}_{\alpha} = a_{\alpha} +b_{\alpha}e^{2r_{0}\tau}\, .  
\end{equation}

\noindent
This ansatz has to be used in the three equations (\ref{eq:1}), (\ref{eq:2})
and (\ref{eq:3}) to determine the actual values of the integration constants
$a_{\alpha},b_{\alpha}$. In the following sections we are going to see how
this ansatz works in particular models, showing that the original differential
equations are solved by the ansatz if the integration constants satisfy
certain algebraic equations that related them to the charges $\mathcal{Q}^{M}$
and non-extremality parameter $r_{0}$, and we will argue that it should always
work, even if the algebraic equations for the integration constants are in
general difficult to solve.

Observe that, since in most cases $e^{-2U_{\rm e}}(H)$ is homogenous of second
degree in the harmonic functions, following the same steps as in the RN
example, we expect to find the event horizon in the $\tau\rightarrow -\infty$
limit and the inner horizon $\tau\rightarrow +\infty$ limit, which will allow
us to find the entropies and temperatures using Eq.~(\ref{eq:2ST}).


\section{Axion-dilaton black holes}
\label{sec-axidilaton}

The so-called axion-dilaton black holes\footnote{For references on these
  black-hole solutions see Refs.~\cite{kn:axidilaton}.} are solutions of the
$\bar{n}=2$ theory with prepotential

\begin{equation}
\mathcal{F}=-i\mathcal{X}^{0}\mathcal{X}^{1}\, .  
\end{equation}

\noindent
This theory has only one complex scalar that it is usually called $\tau$ but
we are going to call $\lambda$ to distinguish it from the radial
coordinate. This scalar is given by

\begin{equation}
\lambda\equiv i\mathcal{X}^{1}/\mathcal{X}^{0}\, .
\end{equation}

\noindent
In terms of $\lambda$ the period matrix is given by

\begin{equation}
(\mathcal{N}_{\Lambda\Sigma})=
\left(
  \begin{array}{cc}
-\lambda & 0 \\
0 & 1/\lambda \\
  \end{array}
\right)  
\end{equation}

\noindent
and, in the $\mathcal{X}^{0}=i/2$ gauge, the K\"ahler potential 
and metric are

\begin{equation}
\mathcal{K}=-\ln{\Im{\rm m}\lambda}\, ,  
\hspace{1cm}
\mathcal{G}_{\lambda\lambda^{*}} = (2\Im{\rm m}\lambda)^{-2}\, .
\end{equation}

\noindent
The reality of the K\"ahler potential requires the positivity of $\Im{\rm
  m}\lambda$. Therefore, $\lambda$ parametrizes the coset
$SL(2,\mathbb{R})/SO(2)$ and the action for the bosonic fields is

\begin{equation}
\label{eq:axidilatonaction}
\begin{array}{rcl}
I
& = & 
{\displaystyle\int} d^{4}x \sqrt{|g|}
\left\{
R 
+{\displaystyle\frac{\partial_{\mu}\lambda \partial^{\mu}\lambda^{*}}{2(\Im {\rm m}\lambda)^{2}}}
-2 \Im{\rm m} \lambda \left[(F^{0})^{2} +|\lambda|^{-2}(F^{1})^{2}\right]
\right.
\\
& & \\
& & 
+2 \Re{\rm e} \lambda \left[F^{0}\star F^{0}  -|\lambda|^{-2}F^{1}\star F^{1}\right]
\biggr \}\, .
\end{array}
\end{equation}

This theory is a truncation of $N=4,d=4$ supergravity. After replacing the
matter vector field $A^{1}$ by its dual ($F_{1}= \Im {\rm m}\lambda \star F^{1}+
\Re{\rm e} \lambda F^{1}$) the action takes the more (manifestly) symmetric form

\begin{equation}
\label{eq:axidilatonaction2}
I
 =  
\int d^{4}x \sqrt{|g|}
\left\{
R 
+{\displaystyle\frac{\partial_{\mu}\lambda \partial^{\mu}\lambda}{(2\Im {\rm m}\lambda)^{2}}}
-2 \Im{\rm m} \lambda \left[(F^{0})^{2} +(F_{1})^{2}\right]
+2 \Re{\rm e} \lambda \left[F^{0}\star F^{0}+F_{1}\star F_{1}\right]
\right\}\, ,
\end{equation}

\noindent
in which it has been exhaustively studied
\cite{Gibbons:1982ih}--\cite{LozanoTellechea:1999my}. In particular, the most
general (non-extremal and rotating) black-holes of this theory were presented
in Ref.~\cite{LozanoTellechea:1999my}. A preliminary check shows that in the
static case the metric and scalars are, in the coordinate $\tau$, of the form
of our deformation ansatz, but we want to reobtain the non-extremal solutions
using the ansatz and the language and notation of $N=2,d=4$ supergravities.

In order to apply the formalism reviewed in the previous section, let us start
by constructing the black-hole potential.

The canonically normalized symplectic section $\mathcal{V}$ is, in a certain
gauge,

\begin{equation}
\mathcal{V}=\frac{1}{2(\Im {\rm m}\lambda)^{1/2}}
\left( 
  \begin{array}{c}
i \\ \lambda \\ -i\lambda \\ 1 \\
  \end{array}
\right)\, ,  
\end{equation}

\noindent
and, in terms of the complex combinations

\begin{equation}
\Gamma_{1} \equiv p^{1} +i q_{0}\, ,
\hspace{1.5cm}  
\Gamma_{0} \equiv q_{1} -i p^{0}\, ,
\end{equation}

\noindent
the central charge and its holomorphic covariant derivative and  the
black-hole potential are

\begin{equation}
  \begin{array}{rcl}
\mathcal{Z}
& = &  
{\displaystyle\frac{1}{2\sqrt{\Im {\rm m}\lambda}}}
\left[\, \Gamma^{*}_{1}-\Gamma_{0}^{*}\lambda\, \right]\, ,
\\
& & \\
\mathcal{D}_{\lambda}\mathcal{Z} 
& = & 
{\displaystyle\frac{i}{4(\Im {\rm m}\lambda)^{3/2}}}
\left[\, \Gamma^{*}_{1}-\Gamma_{0}^{*}\lambda^{*}\, \right]\, ,
\\
& & \\
-V_{\rm bh}
& = & 
{\displaystyle\frac{1}{2 \Im {\rm m}\lambda}}
\left[\, |\Gamma_{1}|^{2}
-2\Re{\rm e}(\Gamma_{1}\Gamma_{0}^{*}) \Re{\rm e} \lambda
+|\Gamma_{0}|^{2}|\lambda|^{2}\, \right]\, .\\
\end{array}
\end{equation}

It is convenient to define the charge

\begin{equation}
\tilde{\mathcal{Z}}
\equiv 
{\displaystyle\frac{1}{2\sqrt{\Im {\rm m}\lambda}}}
\left[\, \Gamma^{*}_{1}-\Gamma_{0}^{*}\lambda^{*}\, \right]\, ,  
\end{equation}

\noindent
in terms of which 

\begin{equation}
\mathcal{G}^{ij^{*}}\mathcal{D}_{i}\mathcal{Z}\mathcal{D}_{j^{*}}\mathcal{Z}^{*} 
=
|\tilde{\mathcal{Z}}|^{2}\, ,
\end{equation}

\noindent
so we can write

\begin{equation}
-V_{\rm bh}= |\mathcal{Z}|^{2}+|\tilde{\mathcal{Z}}|^{2}\, ,
\hspace{1cm}
-\partial_{\lambda}V_{\rm bh}
= 
2\mathcal{Z}^{*}\mathcal{D}_{\lambda}\mathcal{Z}
\sim 
\mathcal{Z}^{*}\tilde{\mathcal{Z}}\, .  
\end{equation}


\subsection{Flow equations}

The potential term can be expanded in the following way:

\begin{equation}
-\left[e^{2U}V_{\rm bh}-r_{0}^{2}\right] 
= 
\Upsilon^{2} + 4\,\mathcal{G}^{\lambda\lambda^{*}}\Psi\Psi^{*} \, ,
\end{equation}

\noindent
where

\begin{eqnarray}
\label{eq:Upsilon}
\Upsilon 
& = &
\frac{e^{U}}{\sqrt{2}}\sqrt{e^{-2U}r_{0}^{2} + |\mathcal{Z}|^{2} 
+|\tilde{\mathcal{Z}}|^{2} + \sqrt{\left(e^{-2U}r_{0}^{2} + |\mathcal{Z}|^{2} 
+|\tilde{\mathcal{Z}}|^{2}\right)^{2} 
-4 |\mathcal{Z}|^{2} |\tilde{\mathcal{Z}}|^2}}\, ,
\\
& & \nonumber \\
\label{eq:Psi}
\Psi 
& = &
i\frac{e^{2U}\mathcal{Z}^{*}\tilde{\mathcal{Z}}}{4\,\Im {\rm m}\lambda\,\Upsilon}\, .
\end{eqnarray}

\noindent
The vector field generated by $(\Upsilon,\Psi,\Psi^{*})$ is conservative or,
in other words, can be written as a gradient of a generalized superpotential
$Y(U,\lambda,\lambda^*)$

\begin{equation}
(\Upsilon,\Psi,\Psi^{*}) 
= 
(\partial_{U} Y, \partial_{\lambda} Y, \partial_{\lambda^{*}} Y)\, ,
\end{equation}

\noindent
if and only if it is irrotational (i.e.~its curl vanishes). This is the case
here, since

\begin{equation}
\partial_{U}\Psi - \partial_{\lambda}\Upsilon = 
\partial_{U}\Psi^{*} - \partial_{\lambda^{*}}\Upsilon 
= \partial_{\lambda}\Psi^{*} - \partial_{\lambda^{*}}\Psi = 0\, ,
\end{equation}

\noindent
which could have been expected on the basis of the results mentioned in
Section~\ref{general}. The explicit form of the generalized superpotential can
be in principle obtained by integrating Eq.~\eqref{eq:Upsilon}, but in
practice this turns out to be very complicated.

The flow equations (\ref{eq:U-Y}, \ref{eq:phi-Y}), in the conventions of
Eq.~\eqref{eq:2}, now take the form:

\begin{eqnarray}
U^{\prime} 
& = &
\Upsilon\, ,\\
& & \nonumber \\
\lambda^{\prime} 
& = &
2\,\mathcal{G}^{\lambda\lambda^{*}}\Psi^{*}\, .
\end{eqnarray}

In the particular case of the Reissner--Nordstr\"om black hole
(cf.~Section~\ref{RNbh}), the first of these equations reduces to the one
derived in \cite{Miller:2006ay} (and the second is not applicable, since there
are no scalars).  For extremal black holes, studied in greater detail below,
one recovers Eq.~(\ref{eq:U-W}, \ref{eq:phi-W}) with either $W = |\mathcal{Z}|$
(the supersymmetric case) or $W = |\tilde{\mathcal{Z}}|$.


\subsection{The extremal case}


\subsubsection{Critical points}

The critical points of the black hole potential are those for which
$\mathcal{Z} = 0$ or $\tilde{\mathcal{Z}} = 0$. They are two isolated points
in moduli space and only the second is supersymmetric. The situation is
summarized in Table~\ref{tab:axidil}.

\begin{table}
  \centering
  \begin{tabular}{||l|c|c|c|c|c|c||}
\hline\hline
& & & & & & \\
Attractor & $\Im{\rm m} \lambda_{\rm h}$ & 
$|\mathcal{Z}_{\rm h}|^{2}$ &
$|\tilde{\mathcal{Z}}_{\rm h}|^{2}$ & $-V_{{\rm bh\, h}}$ & $M$ & $\Sigma^{\lambda}$ 
\\ 
& & & & & & \\
\hline\hline 
& & & & & & \\
$\lambda^{\rm susy}_{\rm h}= \Gamma_{1}/\Gamma_{0}$ & 
$\Im {\rm  m}(\Gamma_{1}\Gamma^{*}_{0})$ & 
$\Im {\rm  m}(\Gamma_{1}\Gamma^{*}_{0})$ & 
$0$ & 
$\Im {\rm  m}(\Gamma_{1}\Gamma^{*}_{0})$ & $|\mathcal{Z}_{\infty}|$  & 
$2ie^{i{\rm Arg}\mathcal{Z}_{\infty}}
\Im{\rm m}\, \lambda_{\infty}\, \mathcal{Z}_{\infty}^{\prime\, *}$ \\
& & & & & & \\
\hline 
& & & & & & \\
$\lambda^{\rm nsusy}_{\rm h}= \Gamma^{*}_{1}/\Gamma^{*}_{0}$ & 
$-\Im {\rm  m}(\Gamma_{1}\Gamma^{*}_{0})$ & 
$0$ & 
$-\Im {\rm  m}(\Gamma_{1}\Gamma^{*}_{0})$ & 
$-\Im {\rm  m}(\Gamma_{1}\Gamma^{*}_{0})$ & $|\tilde{\mathcal{Z}}_{\infty}|$ & 
$2ie^{-i{\rm Arg}\tilde{\mathcal{Z}}_{\infty}}
\Im{\rm m}\, \lambda_{\infty}\, \mathcal{Z}_{\infty}$ \\
& & & & & & \\
\hline\hline
  \end{tabular}
  \caption{Critical points of the axidilaton model. Here
    we are using the notation $\mathcal{Z}_{\rm h}\equiv
    \mathcal{Z}(\lambda_{\rm h},\lambda^{*}_{\rm h},\mathcal{Q})$ etc. 
    In the supersymmetric case the mass $M$ can be found in the explicit 
    solution or from the saturation of the supersymmetric bound. Then, the scalar 
    charge $\Sigma^{\lambda}$ follows from the general extremality bound (or from 
    the knowledge of the explicit solution). In the non-supersymmetric case we do 
    not have analogous arguments and we need the explicit solution, given in 
    Section~\ref{sec-extremallimitsaxidil}.}
  \label{tab:axidil}
\end{table}

As already said in Section \ref{susysolutions}, the supersymmetric stationary
points of the black hole potential must be a minimum. Indeed, the Hessian
evaluated this point in the real basis has the double eigenvalue

\begin{equation}
\frac{|\Gamma_{0}|^{4}}{2\Im {\rm  m}(\Gamma_{1}\Gamma^{*}_{0})} 
= 
\left.
|\Gamma_{0}|^{2}\mathcal{G}_{\lambda\lambda^{*}}
\right\rvert^{\rm susy}_{\rm h} 
= \frac{(p^{0})^{2} + (q_{1})^{2}}{2(p^{0} p^{1} + q_{0} q_{1})}\, .
\end{equation}

Again referring to Section \ref{susysolutions}, one can expect also the
non-supersymmetric extremal stationary point of our model to be stable (up to
possible flat directions). This is confirmed by the direct calculation of the
Hessian, which has the double eigenvalue

\begin{equation}
-\frac{|\Gamma_{0}|^{4}}{2\Im {\rm  m}(\Gamma_{1}\Gamma^{*}_{0})} 
= 
\left.
|\Gamma_{0}|^{2}\mathcal{G}_{\lambda\lambda^{*}}
\right\rvert^{\rm nsusy}_{\rm h} 
= 
-\frac{(p^{0})^{2} + (q_{1})^{2}}{2(p^{0} p^{1} + q_{0} q_{1})}\, .
\end{equation}

Observe that the supersymmetric stationary point and the non-supersymmetric
extremal stationary point exist for mutually exclusive choices of charges and
that in this example, given that $\tilde{\mathcal{Z}}$ differs from
$\mathcal{Z}$ by complex conjugation in the numerator, one could have also
used, with appropriate modifications, the general supersymmetric argument
\cite{Ferrara:1997tw} to study the stability of the non-supersymmetric
critical point.


\subsubsection{Supersymmetric solutions}

According to the general procedure, the supersymmetric solutions are built out
of the four harmonic functions

\begin{equation}
\mathcal{I}^{M} =  \mathcal{I}^{M}_{\infty} -
\frac{\mathcal{Q}^{M}}{\sqrt{2}}\tau\, . 
\end{equation}

\noindent
In this theory the stabilization equations can be easily solved and
they lead to 

\begin{equation}
\mathcal{R}= 
\left(
  \begin{array}{cc}
0 & -\sigma^{1} \\
\sigma^{1} & 0 \\
  \end{array}
\right)\mathcal{I}\, ,
\end{equation}

\noindent
where $\sigma^{1}$ is the standard Pauli matrix, so

\begin{equation}
e^{-2U}
=\langle\, \mathcal{R} \mid \mathcal{I}\, \rangle =
2(\mathcal{I}^{0}\mathcal{I}^{1}+\mathcal{I}_{0}\mathcal{I}_{1})\, ,
\hspace{1cm}
\lambda = i\frac{\mathcal{L}^{1}/X}{\mathcal{L}^{0}/X}= 
\frac{\mathcal{I}^{1}+i\mathcal{I}_{0}}{\mathcal{I}_{1}-i\mathcal{I}^{0}}\, .  
\end{equation}

\noindent
It is useful to define the complex harmonic functions

\begin{equation}
\label{eq:complexharmonicaxdil}
\mathcal{H}_{1} \equiv \mathcal{I}^{1} +i \mathcal{I}_{0}= 
\mathcal{H}_{1\, \infty} -\frac{\Gamma_{1}}{\sqrt{2}}\tau\, ,
\hspace{1.5cm}  
\mathcal{H}_{0} \equiv \mathcal{I}_{1} -i \mathcal{I}^{0}
=
\mathcal{H}_{0\, \infty} -\frac{\Gamma_{0}}{\sqrt{2}}\tau\, ,
\end{equation}

\noindent
in terms of which we have 

\begin{equation}
e^{-2U}
= 2\Im{\rm m}(\mathcal{H}_{1}\mathcal{H}^{*}_{0})\, ,
\hspace{1cm}
\lambda = \frac{\mathcal{H}_{1}}{\mathcal{H}_{0}}\, .
\end{equation}

\noindent
The solution depends on the charges $\mathcal{Q}$ and on the two complex
constants $\mathcal{H}_{1\, \infty}$ and $\mathcal{H}_{0\, \infty}$. A
combination of them ($\mathcal{H}_{1\, \infty}/\mathcal{H}_{0\, \infty}$) is
$\lambda_{\infty}$ and the other combination is determined in terms of
$\mathcal{Q}$ and $\lambda_{\infty}$ by imposing asymptotic flatness

\begin{equation}
2\Im{\rm m}(\mathcal{H}_{1\, \infty}\mathcal{H}^{*}_{0\, \infty})=1\, ,
\end{equation}

\noindent
which provides one real condition, and absence of NUT charge 

\begin{equation}
\Re{\rm e}(\mathcal{H}_{1\, \infty}\Gamma^{*}_{0}
-\mathcal{H}_{0\,\infty}\Gamma^{*}_{1})=0\, , 
\end{equation}

\noindent
which is another real condition. These conditions have two solutions

\begin{equation}
\mathcal{H}_{1\, \infty} = \lambda_{\infty} \mathcal{H}_{0\, \infty}\, , 
\hspace{1.5cm}
\mathcal{H}_{0\, \infty} = 
\mp \frac{i}{\sqrt{2 \Im{\rm m}\lambda_{\infty}}} 
\frac{\mathcal{Z}^{*}_{\infty}}{|\mathcal{Z}_{\infty}|}\, ,  
\hspace{1.5cm}
\mathcal{Z}_{\infty}\equiv
\mathcal{Z}(\lambda_{\infty},\lambda^{*}_{\infty},\mathcal{Q})\, ,
\end{equation}

\noindent
but, using them in the expression for the mass

\begin{equation}
M = \tfrac{1}{\sqrt{2}}\Im{\rm m}(\mathcal{H}_{1\, \infty}\Gamma^{*}_{0}
-\mathcal{H}_{0\,\infty}\Gamma^{*}_{1})\, ,  
\end{equation}

\noindent
one finds that only the upper sign gives a positive mass, which turns out to
be equal to $|\mathcal{Z}_{\infty}|$, as expected. 

The complete solution is, therefore, given by the two harmonic functions

\begin{equation}
\label{eq:Haxidilsusy}
\mathcal{H}^{\rm susy}_{1} 
= 
-\frac{i\lambda_{\infty}}{\sqrt{2 \Im{\rm m}\lambda_{\infty}}} 
\frac{\mathcal{Z}^{*}_{\infty}}{|\mathcal{Z}_{\infty}|}
 -\frac{\Gamma_{1}}{\sqrt{2}}\tau\, ,
\hspace{1.5cm}  
\mathcal{H}^{\rm susy}_{0} 
=
-\frac{i}{\sqrt{2 \Im{\rm m}\lambda_{\infty}}} 
\frac{\mathcal{Z}^{*}_{\infty}}{|\mathcal{Z}_{\infty}|}
-\frac{\Gamma_{0}}{\sqrt{2}}\tau\, .
\end{equation}


\subsubsection{Extremal non-supersymmetric solutions}

According to the proposal made for the $STU$ model in
Ref.~\cite{Kallosh:2006ib}, the metric and scalar fields of the extremal
non-supersymmetric solutions can be constructed by replacing the electric and
magnetic charges of their attractor values by the harmonic functions that have
those charges as coefficients, that is $\mathcal{Q}^{M}$ should be replaced by
the real harmonic function

\begin{equation}
H^{M} = H^{M}_{\infty} - \tfrac{1}{\sqrt{2}} \mathcal{Q}^{M}\tau\, .
\end{equation}

\noindent
The constant parts of the harmonic functions cannot be the same as those of the
supersymmetric solution, otherwise the prescription would lead to

\begin{equation}
e^{-2U}= - 2(\mathcal{I}^{0}\mathcal{I}^{1}+\mathcal{I}_{0}\mathcal{I}_{1})\, ,
\hspace{1cm}
\lambda = 
\frac{\mathcal{I}^{1}-i\mathcal{I}_{0}}{\mathcal{I}_{1}+i\mathcal{I}^{0}}\, ,
\end{equation}

\noindent
or, in terms of the complex harmonic functions defined in
Eq.~(\ref{eq:complexharmonicaxdil})

\begin{equation}
e^{-2U}
= -2\Im{\rm m}(\mathcal{H}_{1}\mathcal{H}^{*}_{0})\, ,
\hspace{1cm}
\lambda = \frac{\mathcal{H}^{*}_{1}}{\mathcal{H}^{*}_{0}}\, .
\end{equation}

If we plug in these expressions the values of the harmonic functions determined
before, we get inconsistent results, because the metric function $e^{-2U}$ is
that of the supersymmetric case and goes to $-1$ at spatial infinity. Thus, the
prescription given in Ref.~\cite{Kallosh:2006ib} should be interpreted as a
replacement of the charges by harmonic functions with asymptotic values yet to
be determined by imposing asymptotic flatness etc.  In
Section~\ref{sec-extremallimitsaxidil} we will determine the form of the
extremal non-supersymmetric solutions by taking an appropriate extremal limit
of the non-extremal solution.


\subsection{Non-extremal solutions}

Our ansatz of Section \ref{generalprescription} for the non-extremal solution is

\begin{equation}
e^{-2U}
= e^{-2[U_{\rm e}(\hat{\mathcal{H}})+r_{0}\tau]}\, ,
\hspace{1cm}
e^{-2U_{\rm e}(\hat{\mathcal{H}})} = 
2\Im{\rm m}(\hat{\mathcal{H}}_{1}\hat{\mathcal{H}}^{*}_{0})\, ,
\hspace{1cm}
\lambda = \lambda_{\rm e}(\hat{\mathcal{H}})= 
\hat{\mathcal{H}_{1}}/\hat{\mathcal{H}_{0}}\, ,
\end{equation}

\noindent
where the deformed harmonic functions are assumed to have the form

\begin{equation}
\hat{\mathcal{H}}_{\Lambda} \equiv A_{\Lambda} +B_{\Lambda}e^{2r_{0}\tau}\, , 
\,\,\,\, \Lambda=1,0\, , 
\end{equation}

\noindent
The four complex constants $A_{\Lambda},B_{\Lambda}$ need to be determined
by imposing on them the equations of motion (\ref{eq:1})--(\ref{eq:3}),
asymptotic flatness, absence of NUT charge plus the definitions of $M$ and
$\lambda_{\infty}$.

Solving the equations of motion is not as complicated a task as it may look at
first sight. First of all, we observe that all the dependence of $U$ and
$\lambda$ on $\tau$ is of the form of the Schwarzschild factor
$e^{2r_{0}\tau}$, which we are going to denote by $f$. Using the chain rule
and combining the first two equations, we get

\begin{eqnarray}
\label{eq:1b}
\ddot{U}_{\rm e} -(\dot{U}_{\rm e})^{2} 
-\mathcal{G}_{ij^{*}}\dot{Z}^{i}  \dot{Z}^{*\, j^{*}} 
& = & 0\, ,\\ 
& & \nonumber \\
\label{eq:2b}
(2r_{0})^{2} \left[f\ddot{U}_{\rm e} +\dot{U}_{\rm e} \right] 
+e^{2U_{\rm e}} V_{\rm bh}
& = & 0\, ,\\
& & \nonumber \\
\label{eq:3b}
(2r_{0})^{2} \left[ f \left(\ddot{Z}^{i}
+\mathcal{G}^{ij^{*}}\partial_{k}\mathcal{G}_{lj^{*}}
\dot{Z}^{k}\dot{Z}^{l}\right) +\dot{Z}^{i} \right]
+e^{2U_{\rm e}}\mathcal{G}^{ij^{*}}\partial_{j^{*}}V_{\rm bh}
& = & 0\, .
\end{eqnarray}

\noindent
Secondly, $U_{\rm e}$ and $\lambda$ only depend on $f$ through the deformed
harmonic functions and, therefore, by virtue of the chain rule:

\begin{equation}
\begin{array}{rcl}
\dot{U}_{\rm e} 
& = & 
\partial_{\Lambda}U_{\rm e}\, B_{\Lambda} +\mathrm{c.c.}\, ,
\\
& & \\   
\ddot{U}_{\rm e} 
& = & 
\partial_{\Sigma}\partial_{\Lambda}U_{\rm e}\, B_{\Lambda}B_{\Sigma} 
+\partial^{*}_{\Sigma}\partial_{\Lambda}U_{\rm e}\, B_{\Lambda}B^{*}_{\Sigma} 
+\mathrm{c.c.}\, ,
\\
& & \\   
\dot{Z}^{i}
& = & 
\partial_{\Lambda}Z^{i}\, B_{\Lambda} +\partial^{*}_{\Lambda}Z^{i}\, B^{*}_{\Lambda}\, ,
\end{array}
\end{equation}

\noindent
etc., where
$\partial_{\Lambda}\equiv \partial/\partial\hat{\mathcal{H}}_{\Lambda}$ and
$\partial^{*}_{\Lambda}\equiv \partial/\partial\hat{\mathcal{H}}^{*}_{\Lambda}$.
Then Eq.~(\ref{eq:1b}) becomes, after multiplication by a convenient global
factor, a quadratic polynomial in the deformed harmonic functions with
coefficients that are combinations of the integration constants
$B_{\Lambda}$. This is true for any $N=2$ model. For the axidilaton model, the
polynomial turns out to be the square of a generalization of the condition of
absence of NUT charge:

\begin{equation}
\Re{\rm e}(\mathcal{H}_{1}B^{*}_{0}
-\mathcal{H}_{0}B^{*}_{1})=
\Re{\rm e}(A_{1}B^{*}_{0}
-A_{0}B^{*}_{1})\, . 
\end{equation}

\noindent
Setting this quantity to zero yields an algebraic equation for the integration
constants, which is enough to solve the first equation. In a similar fashion we
find that the other two differential equations are solved by our ansatz if the
integration constants satisfy certain algebraic constraints that we summarize
here:

\begin{eqnarray}
\Re{\rm e}(A_{1}B^{*}_{0} -A_{0}B^{*}_{1})  
& = & 
0\, ,
\\
& & \nonumber \\
|\Gamma_{1}|^{2}A_{0}B_{0}
+|\Gamma_{0}|^{2}A_{1}B_{1}
-\Re{\rm e}(\Gamma_{1}\Gamma^{*}_{0})(A_{1}B_{0}+A_{0}B_{1})
& = & 
0\, ,
\\
& & \nonumber \\
|\Gamma_{1}|^{2}A^{2}_{0}
+|\Gamma_{0}|^{2}A^{2}_{1}
-2\Re{\rm e}(\Gamma_{1}\Gamma^{*}_{0})A_{1}A_{0}
+8ir_{0}^{2} \Im{\rm m}(A_{1}A^{*}_{0})(A_{1}B_{0}-A_{0}B_{1})
& = & 
0\, ,
\\
& & \nonumber \\
|\Gamma_{1}|^{2}B^{2}_{0}
+|\Gamma_{0}|^{2}B^{2}_{1}
-2\Re{\rm e}(\Gamma_{1}\Gamma^{*}_{0})B_{1}B_{0}
-8ir_{0}^{2} \Im{\rm m}(B_{1}B^{*}_{0})(A_{1}B_{0}-A_{0}B_{1})
& = & 
0\, ,
\\
& & \nonumber \\
\Re{\rm e}(A_{0}B^{*}_{0}) + \frac{1}{8r_{0}^{2}}|\Gamma_{0}|^{2}
& = & 
0\, ,
\\
& & \nonumber \\
\Re{\rm e}(A_{1}B^{*}_{1}) + \frac{1}{8r_{0}^{2}}|\Gamma_{1}|^{2}
& = & 
0\, ,
\\
& & \nonumber \\
\Re{\rm e}(A_{0}B^{*}_{1}+A_{1}B^{*}_{0}) + \frac{1}{4r_{0}^{2}}
\Re{\rm e}(\Gamma_{1}\Gamma_{0}^{*})
& = & 
0\, ,
\end{eqnarray}

\noindent
and to which we must add the conditions of asymptotic flatness and the definitions
of $M$ and $\lambda_{\infty}$:

\begin{eqnarray}
2\Im{\rm m}[(A_{1}+B_{1})(A^{*}_{0}+B^{*}_{0})]
& = & 1\, ,\\
& & \nonumber \\
2r_{0}\Im{\rm m}[A_{1}A^{*}_{0}-B_{1}B^{*}_{0}]
& = & 
M\, ,\\
& & \nonumber \\
\frac{A_{1}+B_{1}}{A_{0}+B_{0}}  
& = & 
\lambda_{\infty}\, .
\end{eqnarray}

From these equations we can derive a relation between the non-extremality
parameter, mass, charge and moduli, which is convenient to write in this form:

\begin{equation}
M^{2} r_{0}^{2} = 
(M^{2} - |\mathcal{Z}_{\infty}|^{2})(M^{2}
-|\tilde{\mathcal{Z}}_{\infty}|^{2})\, .  
\end{equation}

This shows that there are two different extremal limits (supersymmetric and
non-supersymmetric) and that the non-extremal family of solutions interpolates
between these two limits. This will allow us to obtain the extremal
non-supersymmetric solution in a clean way. Observe that in the context of
$N=4$ supergravity both extremal limits are supersymmetric
\cite{Kallosh:1992ii,LozanoTellechea:1999my}.

Expanding the above expression and comparing with the general result
Eq.~(\ref{eq:generalbound}) one can find the scalar charge up to a phase. From
the complete solution (see later) we obtain the exact result

\begin{equation}
\Sigma^{\lambda} 
=  
\frac{2i\Im{\rm m} \lambda_{\infty}\, 
\mathcal{Z}_{\infty} \mathcal{Z}^{\prime\, *}_{\infty}}{M}\, .  
\end{equation}

\noindent
Since the expressions for the metric function and the scalar are invariant if
we multiply $\mathcal{H}_{1},\mathcal{H}_{0}$ by the same phase, we can use
this freedom to simplify the equations setting $\Im{\rm m}(A_{0}+B_{0})=0$. We
can later restore the phase by studying the supersymmetric extremal
limit.

Under this assumption we find (we use a tilde to stress the fact that these
are not the final values of the integration constants):

\begin{eqnarray}
\tilde{A}_{1} 
& = & 
\frac{\lambda_{\infty}}{2\sqrt{2\Im{\rm m}\lambda_{\infty}}}
\left\{
1
+
\frac{1}{Mr_{0}}
\left\{ M^{2}  +\tfrac{1}{2}V_{{\rm bh}\, \infty}
+\tfrac{i}{2}\left[\frac{1}{\lambda_{\infty}}|\Gamma_{1}|^{2}
-\Re{\rm e}(\Gamma_{1}\Gamma_{0}^{*}) \right]
\right\} \right\}\, ,
\\  
& & \nonumber \\
\tilde{B}_{1} 
& = & 
\frac{\lambda_{\infty}}{2\sqrt{2\Im{\rm m}\lambda_{\infty}}}
\left\{
1
-
\frac{1}{Mr_{0}}
\left\{ M^{2}  +\tfrac{1}{2}V_{{\rm bh}\, \infty}
+\tfrac{i}{2}\left[\frac{1}{\lambda_{\infty}}|\Gamma_{1}|^{2}
-\Re{\rm e}(\Gamma_{1}\Gamma_{0}^{*}) \right]
\right\} \right\}\, ,
\\  
& & \nonumber \\
\tilde{A}_{0} 
& = & 
\frac{1}{2\sqrt{2\Im{\rm m}\lambda_{\infty}}}
\left\{ 
1
+
\frac{1}{Mr_{0}}
\left\{ 
M^{2}  +\tfrac{1}{2}V_{{\rm bh}\, \infty}
-\tfrac{i}{2}\left[ \lambda_{\infty}|\Gamma_{0}|^{2}
-\Re{\rm e}(\Gamma_{1}\Gamma_{0}^{*}) \right]
\right\} \right\}\, ,
\\
& & \nonumber \\
\tilde{B}_{0} 
& = & 
\frac{1}{2\sqrt{2\Im{\rm m}\lambda_{\infty}}}
\left\{ 
1
-
\frac{1}{Mr_{0}}
\left\{ 
M^{2}  +\tfrac{1}{2}V_{{\rm bh}\, \infty}
-\tfrac{i}{2}\left[ \lambda_{\infty}|\Gamma_{0}|^{2}
-\Re{\rm e}(\Gamma_{1}\Gamma_{0}^{*}) \right]
\right\} \right\}\, ,
\end{eqnarray}

\noindent
where we are using the shorthand notation $V_{{\rm bh}\, \infty} \equiv V_{\rm
  bh}(\lambda_{\infty},\lambda_{\infty}^{*},\mathcal{Q})$.

Then, the metric function can be put in the two alternative forms

\begin{equation}
\label{eq:alternativemetrics}
e^{-2U} = 1 \pm \frac{M}{r_{0}}(1-e^{\pm 2r_{0} \tau}) 
+\frac{S_{\pm}}{\pi} \frac{\mathrm{sinh}^{2}r_{0}\tau}{r^{2}_{0}}\, ,   
\end{equation}

\noindent
where $S_{\pm}$ are the entropies associated to the outer ($+$) and
inner ($-$) horizons, given in
Eqs.~(\ref{eq:entropiesaxidil1})--(\ref{eq:entropiesaxidil2}). In any
of these two forms $e^{-2U}$ is a sum of manifestly positive terms
when $r^{2}_{0}>2$ and $S_{\pm}>0$, so all the singularities will be
covered by the horizons when they exist. The conditions under which
this happens will be studied later.


\subsubsection{Supersymmetric and non-supersymmetric extremal limits}
\label{sec-extremallimitsaxidil}

The hatted functions have the following extremal limits ($r_{0}\rightarrow 0$):

\begin{enumerate}
\item The supersymmetric extremal limit, when $M\rightarrow
  |\mathcal{Z}_{\infty}|$

\begin{equation}
\hat{\mathcal{H}}_{1,0}
\stackrel{M\rightarrow  |\mathcal{Z}_{\infty}|}{\longrightarrow}
\,\,\,i\frac{\mathcal{Z}_{\infty}}{|\mathcal{Z}_{\infty}|}\, 
\mathcal{H}^{\rm susy}_{1,0}\, ,
\end{equation}

\noindent
with $\mathcal{H}_{1,2}^{\rm susy}$ given in Eq.~(\ref{eq:Haxidilsusy}).

\item The non-supersymmetric extremal limit, when $M\rightarrow
  |\tilde{\mathcal{Z}}_{\infty}|$

\begin{equation}
\hat{\mathcal{H}}_{1,0}
\stackrel{M\rightarrow  |\tilde{\mathcal{Z}}_{\infty}|}{\longrightarrow}
\,\,\, i\frac{\mathcal{Z}^{\prime\, *}_{\infty}}{|\tilde{\mathcal{Z}}_{\infty}|}\, 
\mathcal{H}^{\rm nsusy}_{1,0}\, ,
\end{equation}

\noindent
with 

\begin{equation}
\label{eq:Haxidilnonsusy}
\mathcal{H}^{\rm nsusy}_{1} 
= 
-\frac{i\lambda_{\infty}}{\sqrt{2 \Im{\rm m}\lambda_{\infty}}} 
\frac{\tilde{\mathcal{Z}}_{\infty}}{|\tilde{\mathcal{Z}}_{\infty}|}
 -\frac{\Gamma_{1}^{*}}{\sqrt{2}}\tau\, ,
\hspace{.5cm}  
\mathcal{H}^{\rm nsusy}_{0} 
=
-\frac{i}{\sqrt{2 \Im{\rm m}\lambda_{\infty}}} 
\frac{\tilde{\mathcal{Z}}_{\infty}}{|\tilde{\mathcal{Z}}_{\infty}|}
-\frac{\Gamma_{0}^{*}}{\sqrt{2}}\tau\, .
\end{equation}

$\mathcal{H}^{\rm nsusy}_{1,0}$ can be obtained by replacing everywhere in
$\mathcal{H}^{\rm susy}_{1,0}$ the complex charges $\Gamma_{0,1}$ by their
complex conjugates $\Gamma^{*}_{0,1}$.

We stress that in this case, the metric function and scalar are still given by

\begin{equation}
e^{-2U} 
= 
2\Im{\rm m}(\mathcal{H}^{\rm nsusy}_{1}\mathcal{H}^{{\rm nsusy}\,
  *}_{0})\, ,
\hspace{1cm}
\lambda
=
\mathcal{H}^{\rm nsusy}_{1}/\mathcal{H}^{\rm nsusy}_{0}\, ,
\end{equation}

\noindent
and it is immediate to check that they lead to the non-supersymmetric
attractor and entropy.

\end{enumerate}


\subsubsection{Physical properties of the non-extremal solutions}

The ``entropies'' (one quarter of the areas) of the outer ($+$) and inner ($-$)
horizon, placed at $\tau=-\infty$ and $\tau=+\infty$, respectively, are given
by

\begin{equation}
\label{eq:entropiesaxidil1}
\frac{S_{\pm}}{\pi} = 
(M^{2}-|\mathcal{Z}_{\infty}|^{2})
\pm
(M^{2}-|\tilde{\mathcal{Z}}_{\infty}|^{2})
\pm 2Mr_{0}\, .
\end{equation}

\noindent
They can also be written in the form

\begin{equation}
\label{eq:generalexpressionentropies}
S_{\pm}= 
\pi \left(
\sqrt{N_{\rm R}} \pm  \sqrt{N_{\rm L}}
\right)^{2}\, ,  
\end{equation}

\noindent
with

\begin{equation}
\label{eq:entropiesaxidil2}
N_{\rm R} \equiv M^{2}-|\mathcal{Z}_{\infty}|^{2}\, ,
\hspace{1cm}
N_{\rm L} \equiv M^{2}-|\tilde{\mathcal{Z}}_{\infty}|^{2}\, ,
\end{equation}

\noindent
so the product of these ``entropies'' is manifestly moduli-independent:

\begin{equation}
S_{+}S_{-} = \pi^{2} (N_{\rm R}-N_{\rm L})^{2} = \pi^{2} \left[\Im {\rm
    m}\left(\Gamma_{1}\Gamma_{0}^{*}\right)\,  \right]^{2}\, .   
\end{equation}

From Ref.~\cite{LozanoTellechea:1999my} we know exactly how these
expressions are modified by the introduction of angular momentum $J\equiv
\alpha M$: the entropies are given by

\begin{equation}
\frac{S_{\pm}}{\pi} = 
(M^{2}-|\mathcal{Z}_{\infty}|^{2})
\pm
(M^{2}-|\tilde{\mathcal{Z}}_{\infty}|^{2})
\pm 2M \sqrt{r^{2}_{0}-\alpha^{2}}\, ,  
\end{equation}

\noindent
and can be put in the suggestive form of
Eq.~(\ref{eq:generalexpressionentropies}) with

\begin{equation}
N_{\rm R,L} \equiv
M^{2}-\tfrac{1}{2}(|\mathcal{Z}_{\infty}|^{2}
+|\tilde{\mathcal{Z}}_{\infty}|^{2})
\pm \tfrac{1}{2}\sqrt{(|\mathcal{Z}_{\infty}|^{2}
-|\tilde{\mathcal{Z}}_{\infty}|^{2})^{2}+4J^{2}} \, .
\end{equation}

\noindent
Again, the product of the two entropies is moduli-independent:

\begin{equation}
S_{+}S_{-} = \pi^{2} (N_{\rm R}-N_{\rm L})^{2} = \pi^{2}\left\{ \left[\Im {\rm
    m}\left(\Gamma_{1}\Gamma_{0}^{*}\right)\,  \right]^{2}+4J^{2} \right\}\, .   
\end{equation}

The temperatures $T_{\pm}$ can be computed from $S_{\pm}$ using
Eq.~(\ref{eq:2ST}).

In the two extremal cases, the scalar takes attractor values on the horizon,
which are independent of its asymptotic value $\lambda_{\infty}$.  In
non-extremal black holes the scalar takes the horizon value

\begin{equation}
\lambda^{\rm ne}_{\rm h}
= 
\frac{\lambda_{\infty}S_{+}/\pi 
+i[|\Gamma_{1}|^{2} -\lambda_{\infty}\Re{\rm e} (\Gamma_{1}\Gamma^{*}_{0})]}
{S_{+}/\pi 
-i[\lambda_{\infty}|\Gamma_{0}|^{2} 
-\Re{\rm e} (\Gamma_{1}\Gamma^{*}_{0})]}\, ,  
\end{equation}

\noindent
which manifestly depends on $\lambda_{\infty}$, from which we conclude that
the attractor mechanism does not work in this case.

We observe that if, in the general non-extremal case, $\lambda_{\infty}$ is
set equal to one of the two attractor values, then $\lambda (\tau)$ is
constant over the space. In other words: the non-extremal deformation of a
double-extremal black hole also has constant scalars and, therefore, has the
metric of the non-extremal Reissner--Nordstr\"om black hole.

In the evaporation of a non-extremal black hole of this theory only $M$
changes, while the charges and $\lambda_{\infty}$ remain
constant.\footnote{There are no particles carrying electric or magnetic charges
  in ungauged $N=2,d=4$ supergravity and there is no perturbative physical
  mechanism that can change the moduli, which are properties characterizing
  the vacuum.} The value of $M$ will decrease until it becomes equal to
$\mathrm{max}(|\mathcal{Z}_{\infty}|, |\tilde{\mathcal{Z}}_{\infty}|)$  This
value depends on the values of the charges and moduli in this way:

\begin{equation}
  |\mathcal{Z}_{\infty}|> |\tilde{\mathcal{Z}}_{\infty}|\,\,\,\,
  \Leftrightarrow  \,\,\, 
  \cos \mathrm{Arg}(\lambda_{\infty}/\lambda^{\rm susy}_{\rm h})
  >
  \cos \mathrm{Arg}(\lambda_{\infty}/\lambda^{\rm nsusy}_{\rm h})\, .
\end{equation}

\noindent
Hence, if the phase of $\lambda_{\infty}$ is closer to that of the
supersymmetric attractor value $\Gamma_{1}/\Gamma_{0}$ than to that of the
non-supersymmetric one $\Gamma^{*}_{1}/\Gamma^{*}_{0}$, the central charge
$|\mathcal{Z}_{\infty}|$ will be larger than $|\tilde{\mathcal{Z}}_{\infty}|$
and the evaporation process will stop at the supersymmetric extremal limit and
vice versa.  However, in this analysis we must take into account that the
imaginary part of $\lambda$ must be positive at any point, which means that
$\Im{\rm m}\, \lambda_{\infty}>0$ and only one of $\lambda^{\rm susy}_{\rm h}$
and $\lambda^{\rm nsusy}_{\rm h}$ will satisfy that condition for a given
choice of electric and magnetic charges. Then, it is easy to see that if
$\Im{\rm m}\, \lambda^{\rm susy}_{\rm h}>0$, for any $\lambda_{\infty}$
satisfying $\Im{\rm m}\, \lambda_{\infty}>0$, the above condition is met and
the endpoint of the evaporation process should be the supersymmetric one and,
if $\Im{\rm m}\, \lambda^{\rm nsusy}_{\rm h}>0$ then the opposite will be true
for any admissible $\lambda_{\infty}$.

We conclude that a family of non-extremal black hole solutions with given
electric and magnetic charges $\mathcal{Q}$ and parametrized by $r_{0}$ is
always attracted to one of the two extremal solutions in the evaporation
process, independently of our choice of $\lambda_{\infty}$. The same will
happen to the non-extremal black holes of the model that we are going to
consider next and which can be regarded as an extension of the axidilaton
model.


\section{Black holes of the $\overline{\mathbb{CP}}^{n}$ model}
\label{sec-CPn}

This model is characterized by the prepotential 

\begin{equation}
\mathcal{F} = 
-\tfrac{i}{4}\eta_{\Lambda\Sigma}\mathcal{X}^{\Lambda}\mathcal{X}^{\Sigma}\, ,  
\hspace{1cm}
(\eta_{\Lambda\Sigma}) = \mathrm{diag}(+-\dotsm -)\, ,
\end{equation}

\noindent
and has $n$ scalars

\begin{equation}
Z^{i} \equiv \mathcal{X}^{i}/\mathcal{X}^{0}\, ,  
\end{equation}

\noindent 
to which we add for convenience $Z^{0}\equiv 1$, so we have

\begin{equation}
(Z^{\Lambda}) 
\equiv 
\left(\mathcal{X}^{\Lambda}/\mathcal{X}^{0}\right) = (1,Z^{i})\, ,
\hspace{1cm}
(Z_{\Lambda}) 
\equiv 
(\eta_{\Lambda\Sigma}Z^{\Sigma})= (1,Z_{i}) =(1,-Z^{i})\, .
\end{equation}

\noindent
This will simplify our notation. Thus, the K\"ahler potential and metric are
given by

\begin{equation}
\mathcal{K} = 
-\log{(Z^{*\Lambda}Z_{\Lambda})}\, ,  
\hspace{.5cm}
\mathcal{G}_{ij^{*}} = 
-e^{\mathcal{K}}\left( \eta_{ij^{*}} 
-e^{\mathcal{K}}Z^{*}_{i}Z_{j^{*}} \right)\, ,
\hspace{.5cm}
\mathcal{G}^{ij^{*}} = 
-e^{-\mathcal{K}} \left(\eta^{ij^{*}} 
+Z^{i}Z^{*\, j^{*}}\right)\, .
\end{equation}

The covariantly holomorphic symplectic section reads

\begin{equation}
\mathcal{V}
=
e^{\mathcal{K}/2}
\left(
  \begin{array}{c}
  Z^{\Lambda} \\ \\ -\tfrac{i}{2} Z_{\Lambda} \\
  \end{array}
\right)\, ,
\end{equation}

\noindent
and, in terms of the complex charge combinations

\begin{equation}
\Gamma_{\Lambda}   
\equiv 
q_{\Lambda} +\tfrac{i}{2}\eta_{\Lambda\Sigma}p^{\Sigma}\, ,
\end{equation}

\noindent
the central charge, its holomorphic K\"ahler-covariant derivative and the
black-hole potential are given by

\begin{equation}
\label{eq:centralchargescpn}
\begin{array}{rcl}
\mathcal{Z} 
& = & 
e^{\mathcal{K}/2}Z^{\Lambda}\Gamma_{\Lambda}\, ,   
\\
& & \\
\mathcal{D}_{i}\mathcal{Z} 
& = &
e^{3\mathcal{K}/2}Z^{*}_{i}Z^{\Lambda}\Gamma_{\Lambda}
-e^{\mathcal{K}/2}\Gamma_{i}\, ,
\\
& & \\
|\tilde{\mathcal{Z}}|^{2}
& \equiv & 
\mathcal{G}^{ij^{*}}\mathcal{D}_{i}\mathcal{Z}
\mathcal{D}_{j^{*}}\mathcal{Z}^{*} 
= 
e^{\mathcal{K}}|Z^{\Lambda}\Gamma_{\Lambda}|^{2}  
-\Gamma^{*\, \Lambda}\Gamma_{\Lambda}\, , 
\\
& & \\
-V_{\rm bh}
& = & 
2e^{\mathcal{K}}|Z^{\Lambda}\Gamma_{\Lambda}|^{2} 
-\Gamma^{*\, \Lambda}\Gamma_{\Lambda}\, .
\end{array}
\end{equation}


\subsection{Flow equations}

Similarly as in the axion-dilaton model, the potential term can be expanded
into

\begin{equation}\label{eq:VUpsilonPsi}
-\left[e^{2U}V_{\rm bh}-r_{0}^{2}\right] 
= \Upsilon^{2} + 4\,\mathcal{G}^{ij^{*}}\Psi_{i}\Psi_{j^{*}}^{*} \, ,
\end{equation}

\noindent
where

\begin{eqnarray}
\Upsilon 
& = &
\frac{e^{U}}{\sqrt{2}}\sqrt{|\mathcal{Z}|^{2} 
+|\tilde{\mathcal{Z}}|^{2} 
+e^{-2U}r_{0}^{2} 
+\sqrt{\left(|\mathcal{Z}|^{2} 
+|\tilde{\mathcal{Z}}|^{2}+e^{-2U}r_{0}^{2}\right)^{2} 
-4|\mathcal{Z}|^2 |\tilde{\mathcal{Z}}|^2}}\, ,
\\
& & \nonumber \\
\Psi_{i} 
& = &
e^{2U}\frac{\mathcal{Z}^{*}\,\mathcal{D}_{i}\mathcal{Z}}{\Upsilon}\, , 
\end{eqnarray}

\noindent
with the definitions Eqs.~\eqref{eq:centralchargescpn}.

Since

\begin{equation}
\partial_{U}\Psi_{i} - \partial_{i}\Upsilon 
= 
\partial_{i}\Psi_{j} - \partial_{j}\Psi_{i} 
= 
\partial_{i*}\Psi_{j} - \partial_{j}\Psi_{i^{*}}^{*} = 0\, ,
\end{equation}

\noindent
there exists a superpotential, whose gradient generates the vector field
$(\Upsilon,\Psi_{i},\Psi_{j^{*}}^{*})$ and the first-order equations

\begin{eqnarray}
U^{\prime} & = & \Upsilon\, ,\\
& & \nonumber \\
Z^{i\, \prime} & = & 2\,\mathcal{G}^{ij^{*}}\Psi_{j^{*}}^{*}
\end{eqnarray}

\noindent
solve the second-order equations of motion.


\subsection{The extremal case}


\subsubsection{Critical points}

To find the critical points of the black-hole potential it is simpler to
search for the zeros of

\begin{equation}
\mathcal{G}^{ij^{*}}\partial_{j^{*}}V_{\rm bh} 
= 
2  Z^{\Lambda}\Gamma_{\Lambda} \left(\Gamma^{*\, i} 
-\Gamma^{*\, 0}Z^{i}\right)\, ,
\end{equation}

\noindent
which has two factors that can vanish separately. The second factor vanishes
only for the isolated point in moduli space

\begin{equation}
Z^{i}_{\rm h} = \Gamma^{*\, i}/\Gamma^{*\, 0}\, ,  
\end{equation}

\noindent
and corresponds to the supersymmetric attractor, whereas the first factor
vanishes for the complex hypersurface of the moduli space defined by the
condition

\begin{equation}
Z^{\Lambda}_{\rm h}\Gamma_{\Lambda} =0\, .
\end{equation}

\noindent
These points are associated with non-supersymmetric black holes (the central
charge vanishes). The attractor behavior fixes only a combination of scalars on
the horizon, but each of them individually still depends on the asymptotic
values $Z^{i}_{\infty}$. The situation is summarized in Table~\ref{tab:cpn}.

\begin{table}
  \centering
  \begin{tabular}{||l|c|c|c|c|c||}
\hline\hline
& & & & & \\
Attractor & $e^{-\mathcal{K}_{\rm h}}$ & 
$|\mathcal{Z}_{\rm h}|^{2}$ &
$|\tilde{\mathcal{Z}}_{\rm h}|^{2}$ & $-V_{\rm bh  h}$ & $M$ 
\\ 
& & & & & \\
\hline\hline 
& & & & & \\
$Z^{i\, {\rm susy}}_{\rm h}= \Gamma^{*\, i}/\Gamma^{*\, 0}$ & 
$\Gamma^{*\, \Lambda}\Gamma_{\Lambda}$ & 
$\Gamma^{*\, \Lambda}\Gamma_{\Lambda}$ & 
$0$ & 
$\Gamma^{*\, \Lambda}\Gamma_{\Lambda}$ & $|\mathcal{Z}_{\infty}|$ \\
& & & & & \\
\hline 
& & & & & \\
$Z^{\Lambda\, {\rm nsusy}}_{\rm h}\Gamma_{\Lambda} =0$ & 
$-\Gamma^{*\, \Lambda}\Gamma_{\Lambda}$ & 
$0$ & 
$-\Gamma^{*\, \Lambda}\Gamma_{\Lambda}$ & 
$-\Gamma^{*\, \Lambda}\Gamma_{\Lambda}$ &  $|\tilde{\mathcal{Z}}_{\infty}|$  \\
& & & & & \\
\hline\hline
  \end{tabular}
  \caption{Critical points of the $\overline{\mathbb{CP}}^{n}$ model.}
  \label{tab:cpn}
\end{table}

As we mentioned earlier, the supersymmetric stationary point must be stable
and, since the $\overline{\mathbb{CP}}^{n}$ model is also based on a
homogeneous manifold, the non-supersymmetric stationary points must be stable
as well, even though, because the stationary locus is a submanifold of complex
codimension $1$, rather than an isolated point, one expects $n-1$ complex flat
directions. In fact the Hessian in the real basis has one double eigenvalue

\begin{equation}
4\frac{\delta^{ij}\Gamma^{*}_{i}\Gamma_{j}}{1 
-\delta_{kl}Z^{*\, k}_{\rm h}Z^{l}_{\rm h}}\, .
\end{equation}

At first it may seem that for sufficiently large values of the scalars on the
horizon the eigenvalue could become negative. The above expression, however,
is proportional to a (multiple) eigenvalue of the scalar metric, hence the
values for which the Hessian becomes negative semi-definite would also render
the scalar metric negative definite and are consequently not physically
admissible.


\subsubsection{Supersymmetric solutions}

The stabilization equations are solved by 

\begin{equation}
\mathcal{R}_{\Lambda} = \tfrac{1}{2}\eta_{\Lambda\Sigma}\mathcal{I}^{\Sigma}\, ,
\hspace{1cm}
\mathcal{R}^{\Lambda} = -2\eta^{\Lambda\Sigma}\mathcal{I}_{\Sigma}\, ,  
\end{equation}

\noindent
so

\begin{equation}
\mathcal{L}^{\Lambda}/X =  \mathcal{R}^{\Lambda} +i\mathcal{I}^{\Lambda}
= -2\eta^{\Lambda\Sigma} (\mathcal{I}_{\Sigma} -\tfrac{i}{2}\eta_{\Sigma\Omega}
\mathcal{I}^{\Omega})\, .
\end{equation}

\noindent
Defining the complex combinations of harmonic functions

\begin{equation}
\mathcal{H}_{\Lambda} 
\equiv   
\mathcal{I}_{\Lambda} +\tfrac{i}{2}\eta_{\Lambda\Sigma}\mathcal{I}^{\Sigma}
\equiv
\mathcal{H}_{\Lambda\, \infty} -\tfrac{1}{\sqrt{2}}\Gamma_{\Lambda}\tau \, ,
\end{equation}

\noindent
where $\mathcal{H}_{\Lambda\, \infty}$ are the values at spatial
infinity, we find that the metric function and scalar fields are given
by

\begin{equation}
e^{-2U}=
2\mathcal{H}^{*\, \Lambda}
\mathcal{H}_{\Lambda}\, ,
\hspace{1cm}
Z^{i}= \frac{\mathcal{L}^{i}/X  }{\mathcal{L}^{0}/X} = 
\frac{\mathcal{H}^{* i}}{\mathcal{H}^{* 0}}\, ,
\end{equation}

\noindent
where we are using $\eta$ to raise and lower the indices of the complex
harmonic functions.

The solution depends on the $\bar{n}$ complex charges
$\Gamma_{\Lambda}$ and on the $n+1$ complex constants
$\mathcal{H}_{\Lambda\, \infty}$. $n$ combinations of them are
determined by the asymptotic values of the $n$ scalars

\begin{equation}
  Z^{i}_{\infty} = \mathcal{H}^{*\, i}_{\infty}/\mathcal{H}^{*\, 0}_{\infty}\, ,  
\end{equation}

\noindent
and the remaining one is determined by the two real conditions of
asymptotic flatness

\begin{equation}
2 \mathcal{H}^{*\, \Lambda}_{\infty}
\mathcal{H}_{\Lambda\, \infty} =1\, , 
\end{equation}

\noindent
and absence of NUT charge

\begin{equation}
\Im \mathrm{m} 
\left(\mathcal{H}^{*\, \Lambda}_{\infty}\Gamma_{\Lambda}\right) 
=0\, .
\end{equation}

\noindent
The result is 

\begin{equation}
\mathcal{H}^{\Lambda}_{\infty}  = 
\pm e^{\mathcal{K}_{\infty}/2} \frac{\mathcal{Z}_{\infty}}
{|\mathcal{Z}_{\infty}|}Z^{*\, \Lambda}_{\infty}\, ,
\end{equation}

\noindent
where $\mathcal{K}_{\infty}$ and $\mathcal{Z}_{\infty}$ are the
asymptotic values of the K\"ahler potential and central charge,
although the positivity of the mass, which is given, as expected, by
$M=|\mathcal{Z}_{\infty}|$ allows only for the upper sign.

The complete supersymmetric solution is, therefore, given by the
$\bar{n}$ complex harmonic functions

\begin{equation}
\label{eq:HSUSYcpn}
\mathcal{H}^{\mathrm{susy}}_{\Lambda}  = e^{\mathcal{K}_{\infty}/2} 
\frac{\mathcal{Z}_{\infty}}{|\mathcal{Z}_{\infty}|}
Z^{*}_{\Lambda\, \infty}-\tfrac{1}{\sqrt{2}}\Gamma_{\Lambda}\tau\, , 
\end{equation}

\noindent
that depend only on the $2n+1$ physical complex parameters
$Z^{i}_{\infty},\Gamma_{\Lambda}$.

In order to find the extremal non-supersymmetric solutions we will first
obtain the general non-extremal ones and then we will take the extremal
non-supersymmetric limit. We will see that this procedure works as in
the axidilaton case because the non-extremal solutions interpolate between the
different extremal limits.


\subsection{Non-extremal solutions}

Our ansatz for the non-extremal solution is again

\begin{equation}
e^{-2U}
= e^{-2[U_{\rm e}(\hat{\mathcal{H}})+r_{0}\tau]}\, ,
\hspace{1cm}
e^{-2U_{\rm e}(\hat{\mathcal{H}})} = 
2\hat{\mathcal{H}}^{*\, \Lambda}\hat{\mathcal{H}}_{\Lambda}\, ,
\hspace{1cm}
Z^{i} = Z^{i}_{\rm e}(\hat{\mathcal{H}})= 
\hat{\mathcal{H}}^{*\, i}/\hat{\mathcal{H}}^{*\, 0}\, ,
\end{equation}

\noindent
where the hatted functions are assumed to have the form

\begin{equation}
\hat{\mathcal{H}}^{\Lambda} \equiv A^{\Lambda} +B^{\Lambda}e^{2r_{0}\tau}\, , 
\,\,\,\, \Lambda=0,\dotsc, n\, . 
\end{equation}

As in the axidilaton model, we have to find the $2\bar{n}$ complex constants
$A_{\Lambda},B_{\Lambda}$ by requiring that we have a solution to the equations
of motion (\ref{eq:1b})--(\ref{eq:3b}). It is not difficult to see that this
happens if the following algebraic conditions are satisfied:

\begin{eqnarray}
\label{eq:noNUTcpn}
\Im \mathrm{m}(B^{*\, \Lambda} A_{\Lambda}) 
& = & 
0\, ,
\\ 
& & \nonumber \\
A^{*\, \Lambda}A^{\Sigma}\xi_{\Lambda\Sigma} 
& = & 0\, ,
\\ 
& & \nonumber \\
(A^{*\, \Lambda}B^{\Sigma}+B^{*\, \Lambda}A^{\Sigma})\xi_{\Lambda\Sigma} 
& = & 0\, ,
\\ 
& & \nonumber \\
B^{*\, \Lambda}B^{\Sigma}\xi_{\Lambda\Sigma} 
& = & 0\, ,
\\ 
& & \nonumber \\
(2r_{0})^{2}(B^{*}_{i}A^{*}_{0}-B^{*}_{0}A^{*}_{i})A^{*\, \Lambda} A_{\Lambda}
+(\Gamma^{*}_{i}A^{*}_{0}-\Gamma^{*}_{0}A^{*}_{i})A^{*\, \Lambda} \Gamma_{\Lambda}
& = & 0\, ,
\\ 
& & \nonumber \\
-(2r_{0})^{2}(B^{*}_{i}A^{*}_{0}-B^{*}_{0}A^{*}_{i})B^{*\, \Lambda} B_{\Lambda}
+(\Gamma^{*}_{i}B^{*}_{0}-\Gamma^{*}_{0}B^{*}_{i})B^{*\, \Lambda} \Gamma_{\Lambda}
& = & 0\, ,
\\ 
& & \nonumber \\
(\Gamma^{*}_{i}A^{*}_{0}-\Gamma^{*}_{0}A^{*}_{i})A^{*\, \Lambda} \Gamma_{\Lambda}
+
(\Gamma^{*}_{i}B^{*}_{0}-\Gamma^{*}_{0}B^{*}_{i})B^{*\, \Lambda} \Gamma_{\Lambda}
& = & 0\, ,
\end{eqnarray}

\noindent
where we have defined

\begin{equation}
\xi_{\Lambda\Sigma} \equiv 
2 \left(\Gamma_{\Lambda}\Gamma^{*}_{\Sigma} +8r^{2}_{0} A_{\Lambda}B^{*}_{\Sigma}\right)  
-\eta_{\Lambda\Sigma}\left(\Gamma^{\Omega}\Gamma^{*}_{\Omega} 
+8r^{2}_{0} A^{\Omega}B^{*}_{\Omega}\right)\, .
\end{equation}

In order to fully identify the constants $A_{\Lambda},B_{\Lambda}$ in terms of
the physical parameters, we must add to the above conditions the requirement of
asymptotic flatness and the definitions of mass $M$ and of the asymptotic
values of the scalars $Z^{i}_{\infty}$:

\begin{eqnarray}
2(A^{*\, \Lambda}+B^{*\, \Lambda}) (A_{\Lambda}+B_{\Lambda})  & = & 1\, ,\\
& & \nonumber \\
4\Re\mathrm{e} [B^{*\, \Lambda}(A_{\Lambda}+B_{\Lambda})] & = & 1-M/r_{0}\, ,\\
& & \nonumber \\
\frac{A^{*\, i}+B^{*\, i}}{A^{*\, 0}+B^{*\, 0}} & = & Z^{i}_{\infty}\, .
\end{eqnarray}

The condition of absence of NUT charge arises naturally as a
consequence of the equations of motion (it is Eq.~(\ref{eq:noNUTcpn})).

To solve these equations we choose $A_{0}+B_{0}$ to be real, as we did
in the axidilaton case. Then, we find the following result:

\begin{eqnarray}
A_{\Lambda}
& = & 
\pm \frac{e^{\mathcal{K}_{\infty}/2}}{2\sqrt{2}}
\left\{
Z^{*}_{\Lambda\, \infty}
\left[1+\frac{(M^{2} 
-e^{\mathcal{K}_{\infty}}|Z^{*\, \Sigma}_{\infty}\Gamma^{*}_{\Sigma}|^{2})}{Mr_{0}} 
\right]
+\frac{\Gamma_{\Lambda}Z^{*\, \Sigma}_{\infty}\Gamma^{*}_{\Sigma}}{Mr_{0}}
\right\}\, ,  
\\
& & \nonumber \\
B_{\Lambda}
& = & 
\pm \frac{e^{\mathcal{K}_{\infty}/2}}{2\sqrt{2}}
\left\{
Z^{*}_{\Lambda\, \infty}
\left[1-\frac{(M^{2} 
-e^{\mathcal{K}_{\infty}}|Z^{*\, \Sigma}_{\infty}\Gamma^{*}_{\Sigma}|^{2})}{Mr_{0}} 
\right]
-\frac{\Gamma_{\Lambda}Z^{*\, \Sigma}_{\infty}\Gamma^{*}_{\Sigma}}{Mr_{0}}
\right\}\, ,
\\
& & \nonumber \\
M^{2}r_{0}^{2}
& = & 
(M^{2}-|\mathcal{Z}_{\infty}|^{2})  
(M^{2}-|\tilde{\mathcal{Z}}_{\infty}|^{2})\, ,  
\end{eqnarray}

\noindent
where $|\tilde{\mathcal{Z}}|^{2}$ is defined in
Eq.~(\ref{eq:centralchargescpn}) and we remind the reader that
$-V_{\rm bh} = |\mathcal{Z}|^{2} +|\tilde{\mathcal{Z}}|^{2}$.

With these values it is easy to see that the metric function $e^{-2U}$
can be put in exactly the same form as in the axidilaton case, given
in Eq.~(\ref{eq:alternativemetrics}) where $r_{0}$ and $S_{\pm}$ are
now those of the present case. This means that the metric is regular in
all the $r^{2}_{0}>0$ cases.


\subsubsection{Supersymmetric and non-supersymmetric extremal limits}

Again, there are two possible extremal limits in which $r_{0} \rightarrow 0$:

\begin{enumerate}
\item Supersymmetric, when $M^{2}\rightarrow |\mathcal{Z}|^{2} =
  e^{\mathcal{K}_{\infty}} |Z^{\Sigma}_{\infty}\Gamma_{\Sigma}|^{2}$. In this
  limit we get

\begin{equation}
\hat{\mathcal{H}}_{\Lambda} 
\stackrel{M\rightarrow |\mathcal{Z}_{\infty}|}{\longrightarrow}
\pm \frac{\mathcal{Z}^{*}_{\infty}}{|\mathcal{Z}_{\infty}|} 
\mathcal{H}^{\mathrm{susy}}_{\Lambda}\, ,
\end{equation}

\noindent
where $\mathcal{H}^{\mathrm{susy}}_{\Lambda}$ is given by
Eq.~(\ref{eq:HSUSYcpn}). This determines the phase of $A_{0}+B_{0}$,
which we set to zero at the beginning for simplicity, making use of
the formal phase invariance of the solution.

\item Non-supersymmetric, when $M^{2}\rightarrow
  |\tilde{\mathcal{Z}}|^{2}= e^{\mathcal{K}_{\infty}}
  |Z^{\Sigma}_{\infty}\Gamma_{\Sigma}|^{2} -\Gamma^{*\,
    \Sigma}\Gamma_{\Sigma}$. In this limit we get 

\begin{equation}
\hat{\mathcal{H}}_{\Lambda} 
\stackrel{M\rightarrow |\tilde{\mathcal{Z}}_{\infty}|}{\longrightarrow}
\pm \frac{e^{\mathcal{K}_{\infty}/2}}{2\sqrt{2}}
\left\{
Z^{*}_{\Lambda\, \infty}
-
\frac{1}{|\tilde{\mathcal{Z}}_{\infty}|} 
\left[ -Z^{*}_{\Lambda\, \infty} \Gamma^{*\, \Sigma}\Gamma_{\Sigma}
+\Gamma_{\Lambda}Z^{*\, \Sigma}_{\infty}\Gamma^{*}_{\Sigma} \right]\tau
\right\}\, .
\end{equation}

In this case we do not have an explicit solution to compare with and
we cannot determine the phase of $A_{0}+B_{0}$. However, the metric
and scalar fields do not depend on that phase and the above harmonic
functions determine them completely. 

It takes little time to see that in this case the entropy is 

\begin{equation}
S= -\pi\Gamma^{*\, \Sigma}\Gamma_{\Sigma}\, ,   
\end{equation}

\noindent
as expected, and that on the event horizon the scalars take the values

\begin{equation}
\label{eq:Zhextremalnonsusy}
Z^{*\, i}_{\rm h}
= 
\frac{\Gamma^{i}Z^{*\, \Lambda}_{\infty}\Gamma^{*}_{\Lambda} 
-Z^{*\, i}_{ \infty} \Gamma^{*\, \Sigma}\Gamma_{\Sigma}
 }{\Gamma^{0}Z^{*\, \Gamma}_{\infty}\Gamma^{*}_{\Gamma} 
-\Gamma^{*\, \Omega}\Gamma_{\Omega}}\, ,  
\end{equation}

\noindent
which depend manifestly on the asymptotic values. It is easy to check
that the horizon values satisfy the condition $Z^{\Lambda}_{\rm
  h}\Gamma_{\Lambda}=0$.

\end{enumerate}

Had we tried to implement the prescription of replacement of charges by
harmonic functions in the extremal non-supersymmetric horizon values, it is
difficult to see how the full solution with the above coefficients in the
harmonic functions could have been recovered.


\subsubsection{Physical properties of the non-extremal solutions}

The entropies of the black-hole solutions of this model can also be put in the
form Eqs.~(\ref{eq:entropiesaxidil1})--(\ref{eq:entropiesaxidil2}), where now
$\mathcal{Z}_{\infty}$ and $\tilde{\mathcal{Z}}_{\infty}$ take the form
corresponding to the present model. In both extremal limits we obtain finite
entropies which are moduli-independent, even though in the extremal
non-supersymmetric limit the values of the scalars on the horizon depend on the
asymptotic boundary conditions according to Eq.~(\ref{eq:Zhextremalnonsusy}).
In the non-extremal case, the product of the entropies of the inner and outer
horizon gives the square of the extremal entropy and, consequently, is
moduli-independent.

Also in this case the non-extremal deformation of the double-extremal
solutions have constant scalars: if the asymptotic values of the scalars in
the general case coincide with their horizon attractor values in the extremal
case, then the scalars are constant and the metric is that of the
Reissner--Nordstr\"om solution.

The endpoint of the evaporation process of the non-extremal black holes of this
model is completely determined by their electric and magnetic charges and is
independent of the choice of asymptotic values $Z^{i}_{\infty}$ for the
scalars. Thus, if $\Gamma^{*\, \Lambda}\Gamma_{\Lambda}>0$, which is the
property that characterizes the supersymmetric attractor, then
$|\mathcal{Z}_{\infty}|>|\tilde{\mathcal{Z}}_{\infty}|$ and the evaporation
process will stop when $M=|\mathcal{Z}_{\infty}|$, the supersymmetric case. The
opposite will be true if $\Gamma^{*\, \Lambda}\Gamma_{\Lambda}<0$.  Again, we
can speak of an attractive behavior in the evaporation process.


\section{D0-D4 black holes}
\label{sec-D0-D4}

In this section we are going to obtain, following the procedure outlined in
Section~\ref{generalprescription}, the non-extremal deformation of the
well-known supersymmetric D0-D4 black hole embedded in the $STU$ model
\cite{Duff:1995sm,Behrndt:1996hu,Bellucci:2008sv}.

We have chosen this particular solution because the non-extremal case is
manageable, yet general enough to be interesting. Furthermore, the well-known
supersymmetric limit has a straightforward microscopic interpretation. This
fact could be useful for obtaining a microscopic interpretation in the
non-extremal case, although this interpretation may be difficult to find, since
for non-extremal black holes we have neither supersymmetry nor attractor
mechanism to protect the solution from the effects of a strong-weak change of
the coupling.

The $STU$ model is defined through the following
prepotential:\footnote{Sometimes it is convenient to use the symmetric tensor
  $d_{ijk}=|\epsilon_{ijk}|$ so $\mathcal{F} =
  \frac{1}{6}d_{ijk}\mathcal{X}^{i}\mathcal{X}^{j}\mathcal{X}^{k}/\mathcal{X}^{0}$.}

\begin{equation}
\label{eq:STUprepotential}
\mathcal{F}=\frac{\mathcal{X}^{1}\mathcal{X}^{2}\mathcal{X}^{3}}{\mathcal{X}^{0}}\, ,
\end{equation}

\noindent
and has three scalars customarily defined as

\begin{equation}
Z^{1}\equiv\frac{\mathcal{X}^{1}}{\mathcal{X}^{0}} \equiv S\, ,
\hspace{.5cm}
Z^{2}\equiv\frac{\mathcal{X}^{2}}{\mathcal{X}^{0}} \equiv T\, ,
\hspace{.5cm}
Z^{3}\equiv\frac{\mathcal{X}^{3}}{\mathcal{X}^{0}} \equiv U\, ,
\end{equation}

\noindent
with K\"ahler potential (in the $\mathcal{X}^{0}=1$ gauge) and metric given by

\begin{equation}
e^{-\mathcal{K}}  = -8\, \Im {\rm m}\, S\, \Im {\rm m}\, T\,  
\Im {\rm m}\, U\,
,
\hspace{1cm}
\mathcal{G}_{ij^{*}} = 
\frac{\delta _{(i)j^{*}}}{4( \Im {\rm m}\, Z^{(i)})^{2}}\, .
\end{equation}

The covariantly holomorphic symplectic section is given by 

\begin{equation}
\mathcal{V}  
=
\left(
  \begin{array}{c}
  \mathcal{L}^{\Lambda}  \\ 
  \mathcal{M}_{\Lambda}  \\ 
  \end{array}
\right)
=
e^{\mathcal{K}/2}
\left(
  \begin{array}{c}
1 \\ Z^{i} \\ -\mathcal{F} \\ 3d_{ijk}Z^{j}Z^{k}    
  \end{array}
\right)
=
\frac{1}{2\sqrt{2}\sqrt{-\Im {\rm m}\, S\, \Im {\rm m}\, T\,  
\Im {\rm m}\, U}}
\left(
  \begin{array}{c}
1 \\ S \\ T \\ U \\ -STU \\ TU \\SU \\ ST \\    
  \end{array}
\right)
\, ,
\end{equation}

\noindent
and therefore, we have

\begin{equation}
  \begin{array}{rcl}
\mathcal{Z} 
& = & 
e^{\mathcal{K}/2} W\, ,
\\
& & \\
\mathcal{D}_{i}\mathcal{Z}
& = & 
{\displaystyle\frac{ie^{\mathcal{K}/2}}{2\Im {\rm m}Z^{(i)}}}W_{(i)}\, ,
\\
& & \\
-V_{\rm bh}
& = & 
e^{\mathcal{K}}
\left\{
|W|^{2}
+\sum_{i=1}^{3}|W_i|^{2}
\right\}\, ,
\end{array}
\end{equation}

\noindent
where

\begin{eqnarray}
W=W(S,T,U,\mathcal{Q})
& \equiv & 
-p^{0}\mathcal{F} -q_{0} 
+\sum_{i=1}^{3}\left(3d_{ijk}p^{i}Z^{j}Z^{k} -q_{i}Z^{i}\right)\, ,
\\
& & \nonumber \\ 
W_{1}
& \equiv & 
W(S^{*},T,U,\mathcal{Q})\, ,
\\
& & \nonumber \\ 
W_{2}
& \equiv &
W(S,T^{*},U,\mathcal{Q})\, ,
\\
& & \nonumber \\ 
W_{3}
& \equiv  &
W(S,T,U^{*},\mathcal{Q})\, .
\end{eqnarray}

The D0-D4 black holes that we are going to consider only have four
non-vanishing charges which, when embedded in the $STU$ model, correspond to
three magnetic charges $p^{i}$, $i=1,\dotsc,3$ from the vector fields in the
three vector multiplets, and the electric charge $q_{0}$ of the graviphoton. In
this case the function $W$ reduces to just

\begin{equation}
W=W(S,T,U,\mathcal{Q})
= 3d_{ijk}p^{i}Z^{j}Z^{k}-q_{0}\, .
\end{equation}

Before we analyze the supersymmetric solution, which
eventually is going to be deformed, we discuss the flow equations.


\subsection{Flow equations}

As in Eq.~\eqref{eq:VUpsilonPsi}, also here it is possible to expand the
potential term into squares of

\begin{eqnarray}
\label{eq:UpsilonSTU}
\Upsilon 
& = & 
\tfrac{1}{4}e^{U}
\left(
\sqrt{e^{-2U}r_{0}^{2}+(\hat{q}_{0})^{2}}
+\sum_{j=1}^{3}\sqrt{e^{-2U}r_{0}^{2}+(\hat{p}^{j})^{2}} 
\right)\, ,
\\
& & \nonumber\\
\Psi_{i} 
& = 
& 
\frac{ie^{U}}{16\,\Im{\rm m} Z^{(i)}}
\left(
\sqrt{e^{-2U}r_{0}^{2}+(\hat{q}_{0})^{2}}
-\sum_{j=1}^{3}(-1)^{\delta_{(i)}^{j}}
\sqrt{e^{-2U}r_{0}^{2}+(\hat{p}^{j})^{2}} 
\right)\, ,
\end{eqnarray}

\noindent
where the (hatted) dressed charges are defined as 

\begin{equation}
\label{eq:dressed}
\hat{p}^{i} 
= 
-4 |p^{(i)}|\mathcal{M}_{(i)} 
= 
\sqrt{2}e^{\mathcal{K}/2}d_{(i)jk}|p^{(i)}|
\Im{\rm m}Z^{j}\Im{\rm m}Z^{k}\, , 
\qquad
\hat{q}_{0} 
= 
4|q_{0}|\mathcal{L}^{0} = \sqrt{2}|q_{0}| e^{\mathcal{K}/2}\, .
\end{equation}

The superpotential can be obtained explicitly by integrating
Eq.~\eqref{eq:UpsilonSTU} with respect to $U$:

\begin{equation}
Y = \Upsilon 
- \frac{r_{0}}{4}
\left[
\ln\left(e^{-U}r_{0}^{2}
+r_{0}\sqrt{e^{-2U}r_{0}^{2}+\hat{q}_{0}^{2}}\right)
+\sum_{j=1}^{3}\ln\left(e^{-U}r_{0}^{2}
+r_{0}\sqrt{e^{-2U}r_{0}^{2}+(\hat{p}^{i})^{2}}\right) 
\right]\, ,
\end{equation}

\noindent
and the first-order flow equations take the form:

\begin{eqnarray}
U^{\prime} 
& = & 
\Upsilon = \partial_{U}Y\, ,
\\
& &\nonumber\\
Z^{i\, \prime} 
& = & 
2\,\mathcal{G}^{ij^{*}}\Psi_{j^{*}}^{*}  = 
2\,\mathcal{G}^{ij^{*}}\partial_{j^{*}}Y\, .
\end{eqnarray}


\subsection{The extremal case}


\subsubsection{Critical points}

We start by computing the derivatives of the black-hole potential:

\begin{equation}
-\partial_{Z^{1}}V_{\rm bh} 
=
\frac{ie^{\mathcal{K}}}{\Im {\rm m}\, Z^{1}} 
\left\{W_{1}W^{*} +W^{*}_{2}W^{*}_{3} \right\}=0\, .  
\end{equation}

\noindent
This equation and the other two that can be obtained by permuting $S$ with $T$
and $U$ we get the system

\begin{equation}
\begin{array}{rcl}
W_{1}W^{*} +W^{*}_{2}W^{*}_{3} & = & 0\, ,\\
& & \\    
W_{2}W^{*} +W^{*}_{1}W^{*}_{3} & = & 0\, ,\\
& & \\    
W_{3}W^{*} +W^{*}_{1}W^{*}_{2} & = & 0\, ,\\
& & \\    
\end{array}
\end{equation}

\noindent 
that  admits three kinds of solutions:

\begin{enumerate}
\item $W\neq 0$ and $W_{i}=0\, \,\,\forall i$. This is the $N=2$
  supersymmetric solution because $W_{i}=0$ implies
  $\mathcal{D}_{i}\mathcal{Z}=0$. It corresponds to an isolated point in
  moduli space.

\item $W_{1}\neq 0$, $W=W_{2}=W_{3}=0$ and the other two permutations of this
  solution. These three isolated points in moduli space are not $N=2$
  supersymmetric but correspond to $N=8$ supersymmetric critical points since
  $W$ and the $W_{i}$'s are associated to the four skew eigenvalues of the
  central charge matrix of $N=8$ supergravity \cite{Ferrara:2006em}. Formally
  they can be obtained from the supersymmetric critical point by taking the
  complex conjugate of one of the complex scalars.

\item $|W|=|W_{i}|\, \,\,\forall i$ and $\mathrm{Arg}\, W=
  \sum_{i=1}^{3}\mathrm{Arg}\, W_{i} -\pi$. These are only 4 real equations
  for the 3 complex scalars and admit a 2-parameter space of solutions which
  are not supersymmetric in either $N=2$ or $N=8$ supergravity
  \cite{Ferrara:2006em}. The values of the scalars on the horizon will depend
  on two real combinations of their asymptotic values. 

\end{enumerate}


\subsubsection{Supersymmetric solutions}

Solving directly the equations $W_{i}=0\, \,\,\forall i$ is complicated, but we
can find the supersymmetric attractor values if we can construct the
supersymmetric solutions by the standard method. This requires solving the
stabilization equations or the attractor equations on the horizon, which is not
straightforward either, but has already been done in
Ref.~\cite{Shmakova:1996nz}.

If $\mathcal{I}^{0}\neq 0$, the scalars and metric function of the
supersymmetric extremal solutions are given in terms of the real harmonic
functions $\mathcal{I}^{M}$ by

\begin{align}
Z^{i} & = 
\frac{\mathcal{I}^{\Lambda}\mathcal{I}_{\Lambda} 
-2 \mathcal{I}^{(i)}\mathcal{I}_{(i)}}{2\mathcal{J}_{i}}
-i\frac{e^{-2U}}{4\mathcal{J}_{(i)}}
\\
&  \nonumber \\
e^{-2U}
& =  
2\, \sqrt{
4\mathcal{I}_{0}\mathcal{I}^{1}\mathcal{I}^{2}\mathcal{I}^{3}
-4\mathcal{I}^{0}\mathcal{I}_{1}\mathcal{I}_{2}\mathcal{I}_{3}
+4\sum_{i<j}\mathcal{I}^{i}\mathcal{I}_{i}\mathcal{I}^{j}\mathcal{I}_{j}
-\left(\mathcal{I}^{\Lambda}\mathcal{I}_{\Lambda}\right)^{2}
}\, ,   
\end{align}

\noindent
where

\begin{equation}
\mathcal{J}_{i}
\equiv
3d_{ijk}\mathcal{I}^{j}\mathcal{I}^{k} -\mathcal{I}_{i}\mathcal{I}^{0}\, .  
\end{equation}

If $\mathcal{I}^{0}= 0$, the metric function $e^{-2U}$ and the scalars $Z^{i}$
are the restriction to $\mathcal{I}^{0}= 0$ of the above expressions.

The harmonic functions have the general form
Eq.~(\ref{eq:generalharmonicfunction}) but, as usual, given the charges
$\mathcal{Q}^{M}$, the asymptotic constants $\mathcal{I}^{M}_{\infty}$ are
restricted by the condition of absence of NUT charge Eq.~(\ref{eq:noNUT}).

The simplest supersymmetric extremal D0-D4 black holes, the ones we are
going to consider, have $\mathcal{I}^{0}=\mathcal{I}_{i}=0$
($\mathcal{I}^{0}=\mathcal{I}_{i}=0$ implies $p^{0}=q_{i}=0$ but not the other
way around). The scalars and metric function take the simple forms

\begin{align}
Z^{i} & = 
-4ie^{2U}\mathcal{I}_{0}\mathcal{I}^{i}\, ,
\\
&  \nonumber \\
e^{-2U}
& =  
4\, \sqrt{
\mathcal{I}_{0}\mathcal{I}^{1}\mathcal{I}^{2}\mathcal{I}^{3}
}\, ,   
\end{align}

\noindent
and the condition of absence of NUT charge Eq.~(\ref{eq:noNUT}) is
automatically satisfied for arbitrary values of the constants
$\mathcal{I}^{M}_{\infty}$.

The regularity of the metric and scalar fields (whose imaginary part must be
strictly positive in these conventions) for all $\tau\in\left(-\infty,0\right)$
implies

\begin{equation}
\mathrm{sign}\, \mathcal{I}_{0\, \infty} = 
\mathrm{sign}\, q_{0}\, , 
\hspace{1cm}
\mathrm{sign}\, \mathcal{I}^{i}_{\infty} = 
\mathrm{sign}\, p^{i}\, ,\,\,\forall i\, , 
\end{equation}

\noindent
and the reality of the metric function and negative definiteness of the
imaginary parts of the scalars imply

\begin{equation}
\mathcal{I}_{0}\mathcal{I}^{i} > 0\, ,
\hspace{1cm}
\mathcal{I}_{0}\mathcal{I}^{1}\mathcal{I}^{2}\mathcal{I}^{3}>0\, ,
\end{equation}

\noindent
which leave us with just two options

\begin{equation}
  \begin{array}{rcl}
\mathcal{I}_{0}, \mathcal{I}^{1}, \mathcal{I}^{2}, \mathcal{I}^{3} >0\, ,\,\,\, 
& q_{0}, p^{1}, p^{2}, p^{3} >0\, ,\,\,\, & 
\mathcal{I}_{0\, \infty}, \mathcal{I}^{1}_{\infty}, \mathcal{I}^{2}_{\infty}, \mathcal{I}^{3}_{\infty} >0\, , 
\\
& & \\
\mathcal{I}_{0}, \mathcal{I}^{1}, \mathcal{I}^{2}, \mathcal{I}^{3} < 0\, ,\,\,\, 
& q_{0}, p^{1}, p^{2}, p^{3} < 0\, ,\,\,\, & 
\mathcal{I}_{0\, \infty}, \mathcal{I}^{1}_{\infty}, \mathcal{I}^{2}_{\infty}, 
\mathcal{I}^{3}_{\infty} <0\, . 
\\
\end{array}
\end{equation}

Therefore, in the supersymmetric solution we have two disconnected
possibilities (in the sense that it is not possible to go from one to the
other continuously without making the metric functions or the scalars
singular).

Imposing asymptotic flatness and absence of NUT charge we find that the four
harmonic functions can be written in terms of the physical parameters in the
form

\begin{equation}
\begin{array}{rclcl}
\mathcal{I}_{0}
& = & 
s_{0}\, 
\left\{ {\displaystyle\frac{1}{4\sqrt{2}\mathcal{L}^{0}_{\infty}}
-\frac{1}{\sqrt{2}}}|q_{0}|\tau\right\}
& = & 
{\displaystyle\frac{s_{0}}{4\sqrt{2}\mathcal{L}^{0}_{\infty}}}\, 
\left( 1 -\hat{q}_{0\, \infty}\tau\right)
\, ,
\\
& & & & \\    
\mathcal{I}^{i}
& = & 
s^{(i)}\, 
\left\{ -{\displaystyle\frac{1}{4\sqrt{2}\mathcal{M}_{(i)\, \infty}}
-\frac{1}{\sqrt{2}}}|p^{(i)}|\tau\right\}
& = & 
-{\displaystyle\frac{s^{(i)}}{4\sqrt{2}\mathcal{M}_{(i)\, \infty}}}\, 
\left( 1 - \hat{p}^{(i)}_{\infty}\tau\right)
\, ,
\\
\end{array}
\end{equation}

\noindent
where $s_{0},s^{i}$ are the signs of the charges $q_{0},p^{i}$ and
$\hat{q}_{0\, \infty},\hat{p}^{i}_{\infty}$ are the asymptotic values of the
dressed charges defined in Eq.~(\ref{eq:dressed}). These are positive by
definition. On the other hand, as previously discussed, the signs
$s_{0},s^{i}$ must be either all positive or all negative in the
supersymmetric case.

Plugging these expressions into the metric function we can compute the entropy
and the mass of the black hole, finding

\begin{equation}
\label{eq:Entropy}
S/\pi
=
|\mathcal{Z}(Z_{\rm h},Z^{*}_{\rm h},\mathcal{Q})|^{2}
=
2\sqrt{q_{0} p^{1} p^{2} p^{3}}  \, ,
\end{equation}

\begin{equation}
\label{eq:Mass1}
M
= 
|\mathcal{Z}(Z_{\infty},Z^{*}_{\infty},\mathcal{Q})| 
= 
\tfrac{1}{4}\left(\hat{q}_{0\, \infty}+\hat{p}^{1}_{\infty}
+\hat{p}^{2}_{\infty}+\hat{p}^{3}_{\infty}\right)\, .
\end{equation}


\subsubsection{Extremal, non-supersymmetric solutions}

According to the discussion of the critical points of the black-hole
potential, we can obtain 3 non-supersymmetric extremal black holes by formally
replacing one of the scalars by its complex conjugate. If we do it for
$Z^{1}$, for instance, we get

\begin{equation}
  \begin{array}{rcl}
\Im {\rm m}\, Z^{1} = 
-4e^{2U}\mathcal{I}_{0}\mathcal{I}^{1} 
&
\,\,\,
\longrightarrow
\,\,\,
&
+4e^{2U}\mathcal{I}_{0}\mathcal{I}^{1}\, ,
\\
& & \\
e^{-2U} = 
4 \sqrt{\mathcal{I}_{0}\mathcal{I}^{1}\mathcal{I}^{2}\mathcal{I}^{3}}
&
\,\,\,
\longrightarrow
\,\,\,
&
4 \sqrt{\mathcal{I}_{0}\mathcal{I}^{1}\mathcal{I}^{2}\mathcal{I}^{3}}\, ,
\\
\end{array}
\end{equation}

\noindent
with $\Im {\rm m}\, Z^{1}$ strictly negative. This transformation is equivalent
to the replacement of $\mathcal{I}^{1}$ by $-\mathcal{I}^{1}$ everywhere. To
take into account these and also further possibilities, we write the extremal
solutions in the form

\begin{equation}
Z^{(i)} = 
-4s_{0}s^{(i)}e^{2U}\mathcal{I}_{0}\mathcal{I}^{(i)}\, ,
\hspace{1cm} 
e^{-2U} = 
4 \sqrt{s_{0}s^{1}s^{2}s^{3}\mathcal{I}_{0}\mathcal{I}^{1}
\mathcal{I}^{2}\mathcal{I}^{3}}\, ,
\end{equation}

\noindent
where $s_{0},s^{i}$ are the signs of the respective harmonic functions (which
coincide with those of the charges and those of the asymptotic constants). The
possible choices and their relation to supersymmetry are given in
Table~\ref{tab:stuextreme}.

\begin{table}
  \centering
  \begin{tabular}{||c|c|c|c||}
    \hline\hline
    $s_{0}$ & $s^{1}$ &   $s^{2}$ &   $s^{3}$ \\
    \hline\hline
    $\pm$ & $\pm$ &   $\pm$ &   $\pm$ \\
    \hline\hline
    $\pm$ & $\mp$ &   $\pm$ &   $\pm$ \\
    \hline
    $\pm$ & $\pm$ &   $\mp$ &   $\pm$ \\
    \hline
    $\pm$ & $\pm$ &   $\pm$ &   $\mp$ \\
    \hline\hline
    $\mp$ & $\pm$ &   $\pm$ &   $\pm$ \\
    \hline
    $\mp$ & $\mp$ &   $\pm$ &   $\pm$ \\
    \hline
    $\mp$ & $\pm$ &   $\mp$ &   $\pm$ \\
    \hline
    $\mp$ & $\pm$ &   $\pm$ &   $\mp$ \\
    \hline\hline
  \end{tabular}
  \caption{Possible sign choices for extremal black holes of the D0-D4
    model. The first two possibilities (first row of the table) correspond to
    the $N=2$ supersymmetric black holes. The six choices in the
    2\textsuperscript{nd}, 3\textsuperscript{rd} and 4\textsuperscript{th}
    rows correspond to the extremal black holes that are not supersymmetric
    in $N=2$ supergravity but are supersymmetric when the theory is embedded 
    in the $N=8$ supergravity. The last 8 choices (4 rows) correspond to extremal
    black holes which are not supersymmetric in any theory.}
  \label{tab:stuextreme}
\end{table}

The entropy of these solutions is given by 

\begin{equation}
\label{eq:extremalentropy}
S/\pi=2\sqrt{s_{0} s^{1} s^{2} s^{3} q_{0} p^{1} p^{2} p^{3}}=
2\sqrt{|q_{0} p^{1} p^{2} p^{3}|}  \, ,
\end{equation}

\noindent
and the mass is still given by Eq.~(\ref{eq:Mass1})

\begin{equation}
\label{eq:extremalmass}
M
= 
\tfrac{1}{4}\left(\hat{q}_{0\, \infty}+\hat{p}^{1}_{\infty}
+\hat{p}^{2}_{\infty}+\hat{p}^{3}_{\infty}\right)\, ,
\end{equation}

\noindent
but it coincides with $|\mathcal{Z}(Z_{\infty},Z^{*}_{\infty},\mathcal{Q})|$
only for the first two choices of signs in Table~\ref{tab:stuextreme}. For the
choices in the rows $i+1=2,3,4$ of the table, the mass equals
$e^{\mathcal{K}/2}|W_{i}|$ (for them
$|\mathcal{Z}(Z_{\infty},Z^{*}_{\infty},\mathcal{Q})|=0$) and for the other
eight combinations of signs the mass is numerically equal to
$4|\mathcal{Z}(Z_{\infty},Z^{*}_{\infty},\mathcal{Q})|$. Thus, for all these
extremal black holes $M>|\mathcal{Z}(Z_{\infty},Z^{*}_{\infty},\mathcal{Q})|$.


\subsection{Non-extremal D0-D4 black hole}

According to the general prescription we describe the non-extremal solution
with four functions $\hat{\mathcal{I}}_{0}, \hat{\mathcal{I}}^{1},
\hat{\mathcal{I}}^{2}, \hat{\mathcal{I}}^{3}$ of $\tau$, which we will denote
collectively by $\mathcal{I}^{\Lambda}$ in this section and which we assume to
be of the general form

\begin{equation}
\label{eq:NonExtremalFunctions}
\hat{\mathcal{I}}^{\Lambda} = a^{\Lambda} + b^{\Lambda} e^{2 r_{0} \tau}\, ,
\end{equation}

\noindent
The metric factor and scalar fields are assumed to take the form

\begin{align}
\label{eq:MetricFactor}
e^{-2 U} 
& = 
e^{-2 [U_{\rm e}+r_{0} \tau]}\, ,
\\
& \nonumber \\
\label{eq:Scalars}
Z^{i} 
& = 
-4ie^{2U_{\rm e}}\hat{\mathcal{I}}_{0} \hat{\mathcal{I}}^{i}\, ,
\end{align}

\noindent
where 

\begin{equation}
e^{-2 U_{\rm e}}= 4\sqrt{\hat{\mathcal{I}}_{0} \hat{\mathcal{I}}^{1} 
\hat{\mathcal{I}}^{2} \hat{\mathcal{I}}^{3}} \, .  
\end{equation}

Observe that the consistency of this ansatz requires that all the functions
$\hat{\mathcal{I}}^{\Lambda}$ must be simultaneously positive or
negative. Furthermore, they must be finite in the interval $\tau \in
(-\infty,0)$, which implies that 

\begin{equation}
\label{eq:finitenesscondition}
\mathrm{sign}\, a^{\Lambda} \neq 
\mathrm{sign}\, b^{\Lambda}\, , 
\hspace{1cm}
|a^{\Lambda}| >   |b^{\Lambda}|\, \,\,\,\, \forall\,  \Lambda\, .
\end{equation}

Plugging this ansatz into the Eqs.~(\ref{eq:1b})--(\ref{eq:3b}) we find that they
are solved if the constants $a^{\Lambda},b^{\Lambda}$ satisfy for each
value of $\Lambda$

\begin{equation}
a^{(\Lambda)}b^{(\Lambda)} = -\frac{(p^{\Lambda})^{2}}{8r_{0}^{2}}\, .
\end{equation}

\noindent
In order to determine all the constants in terms of the physical parameters we
impose asymptotic flatness and use the definitions of mass and the asymptotic
values of the scalars, which yield the additional relations (the
condition of absence of NUT charge is automatically satisfied)

\begin{align}
\prod_{\Lambda} (a^{\Lambda}+b^{\Lambda}) 
& = 
\frac{1}{16}\, ,
\\ 
& \nonumber \\
\sum_{\Lambda} \frac{b^{\Lambda}}{a^{\Lambda}+b^{\Lambda}} 
& = 
2
\left(
1-\frac{M}{r_{0}}
\right)\, ,
\\
& \nonumber \\
\Im {\rm m}\,  Z^{i}_{\infty}
& = 
-4(a_{0}+b_{0})(a^{i} +b^{i})\, .
\end{align}

The solution to these equations that satisfies the finiteness condition
Eq.~(\ref{eq:finitenesscondition}) is

\begin{align}
\left(
\begin{array}{c}
a_{0} \\ b_{0} \\    
\end{array}
\right)  
& =  
\frac{\varepsilon}{8\sqrt{2} \mathcal{L}^{0}_{\infty}}
\left\{
1
\pm 
\frac{1}{r_{0}}
\sqrt{r_{0}^{2} + (\hat{q}_{0\, \infty})^{2}}
\right\}\, ,
\\
 & \nonumber \\
\left(
\begin{array}{c}
a^{i} \\ b^{i} \\    
\end{array}
\right)  
& =  
-\frac{\varepsilon}{8\sqrt{2} \mathcal{M}_{i\, ,\infty}}
\left\{
1
\pm 
\frac{1}{r_{0}}
\sqrt{r_{0}^{2} + (\hat{p}^{i}_{\infty})^{2}}
\right\}\, ,
\end{align}

\noindent
where the upper sign corresponds to the constant $a$ and the lower to $b$ and
$\varepsilon$ is the global sign of the functions
$\hat{\mathcal{I}}^{\Lambda}$. We must stress that, unlike in the
extremal case, this sign is not related to that of the charges.


\subsubsection{Physical properties of the non-extremal solutions}

The mass is given by 

\begin{equation}
\label{eq:nonextremalmass}
M =\tfrac{1}{4} \sum_{\Lambda}\sqrt{r_{0}^{2}
  +(\hat{p}^{\Lambda}_{\infty})^{2}}\, ,  
\end{equation}

\noindent
and it is evident that in the extremal limit it takes the value
Eq.~(\ref{eq:extremalmass}), while the entropies are given by 

\begin{equation}
\label{eq:entropy+and-}
\frac{S_{\pm}}{\pi}
=
\frac{A_{\pm}}{4\pi}=  
\sqrt{\prod_{\Lambda}
\left(
r^{2}_{0} \pm \sqrt{r^{2}_{0} + \left(\hat{p}^{\Lambda}_{\infty}\right)^{2}}
\right)}
\, ,
\end{equation}

\noindent
and take the value Eq.~(\ref{eq:extremalentropy}), since $\prod_{\Lambda}
|\hat{p}^{\Lambda}_{\infty}|=\prod_{\Lambda} |p^{\Lambda}|$. Observe that

\begin{equation}
\label{eq:entropy+-}
\frac{S_{+}}{\pi}\frac{S_{-}}{\pi}= 4 |q_{0} p^{1} p^{2} p^{3}|\, ,
\end{equation}
 
\noindent
which is the square of the moduli-independent entropy of all the extremal
black holes.

It is highly desirable to have an explicit expression of the non-extremality
parameter $r_{0}$ in terms of the physical parameters
$M,p^{\Lambda},Z^{i}_{\infty}$, which, in turn, would allow us to express mass
and entropy as functions of $p^{\Lambda},Z^{i}_{\infty}$ alone. Furthermore,
such an expression would allow us to study the different extremal limits or
relations between $M$ and $p^{\Lambda}$ and $Z^{i}_{\infty}$ that make $r_{0}$
vanish. In the general case, solving Eq.~(\ref{eq:nonextremalmass}) explicitly
is impossible, though.  We can, nevertheless, consider some particular
examples, obtained by fixing the relative values of the dressed charges
$\hat{p}^{i}$ and $\hat{q}_{0}$:

\begin{enumerate}

\item If $\hat{p}^{1}_{\infty}=\hat{p}^{2}_{\infty} =\hat{p}^{3}_{\infty}
  =\hat{q}_{0\, \infty}$, then Eq.~(\ref{eq:nonextremalmass}) simplifies to:

\begin{equation}
\label{eq:MasaNoExtremal1}
M=\sqrt{r^{2}_{0}+\left(\hat{q}_{0\, \infty}\right)^{2}}\, ,
\end{equation}
 
\noindent
so

\begin{equation}
\label{eq:MasaNoExtremal1Factor}
r^{2}_{0}=\left(M-\hat{q}_{0\, \infty}\right)
\left(M+\hat{q}_{0\, \infty}\right)\, ,
\end{equation}

\noindent
from which we conclude that we can reach the extremal limit $M=\hat{q}_{0\,
  \infty}$ in two different ways:\footnote{We remind the reader that we have
  defined the dressed charges to always be positive.} $M= s_{0}\hat{q}_{0\,
\infty}$ and $M= -s_{0}\hat{q}_{0\, \infty}$. Which one is reached depends on
$s_{0} = \mathrm{sign}\,q_{0}$. Whether this limit is supersymmetric or not will
depend on the signs of the charges, as discussed in Table~\ref{tab:stuextreme}.

We can use Eq.~(\ref{eq:MasaNoExtremal1Factor}) to express the entropy in
terms of the mass, the charges, and the asymptotic values of the scalars at
infinity in the familiar form:

\begin{equation}
\label{eq:entropy1+and-}
\frac{S_{\pm}}{\pi}=\left(\sqrt{N^{(1)}_{\rm R}}\pm \sqrt{N^{(1)}_{\rm L}}\right)^{2}\, ,
\end{equation}

\noindent
where 

\begin{equation}
N^{(1)}_{\rm R}=M^{2}\, ,
\hspace{1cm}
N^{(2)}_{\rm L}=M^{2}-\hat{q}_{0\, \infty}^{2}\, . 
\end{equation}


\item If $\hat{p}^{1}_{\infty}=\hat{p}^{2}_{\infty}$ and
  $\hat{p}^{3}_{\infty}=\hat{q}_{0\, \infty}$, then the mass of the black hole
  is given by

\begin{equation}
\label{eq:MasaNoExtremal2}
M= \tfrac{1}{2}\left[
\sqrt{r^{2}_{0}+\left(\hat{p}^{1}_{\infty}\right)^{2}} 
+
\sqrt{r^{2}_{0}+\left(\hat{q}_{0\, \infty}\right)^{2}}\right]\, ,
\end{equation}

\noindent
and Eq.~(\ref{eq:MasaNoExtremal2}) can be inverted to obtain

\begin{equation}
\label{eq:MasaNoExtremal2Factor}
  M^{2} r^{2}_{0} = 
  \left(M^{2}-\frac{(\hat{p}^{1}_{\infty}+\hat{q}_{0\, \infty})^{2}}{4}\right)
\left(M^{2}-\frac{(\hat{p}^{1}_{\infty}-\hat{q}_{0\, \infty})^{2}}{4}\right)\, ,
\end{equation}

\noindent
from which we find four possible extremal limits: 

\begin{equation}
M= 
\left\{
  \begin{array}{c}
\tfrac{1}{2}(s^{1}\hat{p}^{1}_{\infty}+s_{0}\hat{q}_{0\, \infty})\, , 
\\ \\
\tfrac{1}{2}(s^{1}\hat{p}^{1}_{\infty}-s_{0}\hat{q}_{0\, \infty})\, , 
\\ \\
-\tfrac{1}{2}(s^{1}\hat{p}^{1}_{\infty}+s_{0}\hat{q}_{0\, \infty})\, , 
\\ \\
-\tfrac{1}{2}(s^{1}\hat{p}^{1}_{\infty}-s_{0}\hat{q}_{0\, \infty})\, .
\end{array}
\right.
\end{equation}

Which extremal limit will be attained if the mass diminishes in the process of
evaporation depends on the signs of the charges $s^{1},s_{0}$ but it will
always be the largest possible value so that

\begin{equation}
M = \tfrac{1}{2}(\hat{p}^{1}_{\infty}+\hat{q}_{0\, \infty})\, .
\end{equation}

In terms of the mass, the charges, and the asymptotic values of the scalars at
infinity, the entropies are again given by 

\begin{equation}
\frac{S_{\pm}}{\pi}=\left(\sqrt{N^{(2)}_{\rm R}}\pm\sqrt{N^{(2)}_{\rm L}}\right)^{2}\, ,
\label{eq:entropy2+-}
\end{equation}

\noindent
where 

\begin{equation}
N^{(2)}_{\rm R}=
M^{2} -
\frac{\left(\hat{p}^{1}_{\infty}+\hat{q}_{0\, \infty}\right)^{2}}{4}\, ,
\hspace{1cm}
N^{(2)}_{\rm L}=
M^{2} -
\frac{\left(\hat{p}^{1}_{\infty}-\hat{q}_{0\, \infty}\right)^{2}}{4}\, ,
\end{equation}

\noindent
and the product of the two entropies gives the moduli-independent entropy of
the extremal black hole with the same charges, squared.


\item If $\hat{p}^{1}_{\infty}=\hat{p}^{2}_{\infty}=\hat{p}^{3}_{\infty}$, then
the mass is given by

\begin{equation}
M = 
\tfrac{1}{4}\left[3\sqrt{r^{2}_{0}+\left(\hat{p}^{1}_{\infty}\right)^{2}} 
+\sqrt{r^{2}_{0}+\left(\hat{q}_{0\, \infty}\right)^{2}}\right]\, ,
\label{eq:MasaNoExtremal3}
\end{equation}

\noindent
This equation can be written in a polynomial form by squaring it several times,
and then it can be solved for $r^{2}_{0}$

\begin{equation}
\label{eq:MasaNoExtremalSolution3r}
 r^{2}_{0} = \tfrac{1}{8} 
\left[
(\hat{q}_{0\, \infty})^{2} - 9 (\hat{p}^{1}_{\infty})^{2} 
+20 M^{2} - 6 \sqrt{2} \sqrt{(\hat{q}_{0\, \infty})^{2} M^{2} 
-(\hat{p}^{1}_{\infty})^{2} M^{2} + 2 M^{4}}\right]\, .
\end{equation}

\noindent
From this equation we can obtain the extremal values of $M$:

\begin{equation}
\label{eq:MasaNoExtremalSolution3}
M= 
\left\{
  \begin{array}{c}
\tfrac{1}{4}(3s^{1}\hat{p}^{1}_{\infty}+s_{0}\hat{q}_{0\, \infty})\, , 
\\ \\
\tfrac{1}{4}(3s^{1}\hat{p}^{1}_{\infty}-s_{0}\hat{q}_{0\, \infty})\, , 
\\ \\
-\tfrac{1}{4}(3s^{1}\hat{p}^{1}_{\infty}+s_{0}\hat{q}_{0\, \infty})\, , 
\\ \\
-\tfrac{1}{4}(3s^{1}\hat{p}^{1}_{\infty}-s_{0}\hat{q}_{0\, \infty})\, .
\end{array}
\right.
\end{equation}

\noindent
The extremal limit that will be reached first in the evaporation process will
be that with the largest value of the mass

\begin{equation}
M= \tfrac{1}{4}(3\hat{p}^{1}_{\infty}+\hat{q}_{0\, \infty})\, , 
\end{equation}

\noindent
and the supersymmetry will depend on the signs of the charges.

As in the previous examples, we can write the entropy in terms of $N^{(3)}_{\rm
  R}$ and $N^{(3)}_{\rm L}$, although in this case the expression for them is
not very manageable. However, we can compute

\begin{equation}
\label{eq:entropy3}
\frac{\sqrt{S_{+} S_{-}}}{\pi}=N^{(3)}_{\rm R}-N^{(3)}_{\rm L}
=  \left(\hat{p}^{1}_{\infty}\right)^{3/2}\sqrt{\hat{q}_{0\, \infty}}
=2\sqrt{|q_{0} p^{1} p^{2} p^{3}|}\, .
\end{equation}

\noindent
Eq.~(\ref{eq:entropy3}) depends only on the charges and it is indeed the
supersymmetric entropy, as already demonstrated in the general case
(\ref{eq:entropy+and-}) and (\ref{eq:entropy+-}).

\end{enumerate} 

In the general case, even though finding a closed-form explicit expression for
$r_{0}\left(Z^{i}_{\infty},\mathcal{Q},M\right)$ is at best unfeasible, it is
still possible to obtain the values $M_{\rm e}=M(Z^{i}_{\infty},\mathcal{Q})$
at which extramality is reached by setting $r_{0}=0$ in
Eq.~(\ref{eq:nonextremalmass}). There are $2^{4}=16$ possible extremal limits
given by

\begin{equation}
M=
\left\{
  \begin{array}{c}
\pm\tfrac{1}{4}
\left(
s_{0}\hat{q}_{0} +s^{1}\hat{p}^{1}+s^{2}\hat{p}^{2}+s^{3}\hat{p}^{3}
\right)\, ,
\\ 
\pm\tfrac{1}{4}
\left(
s_{0}\hat{q}_{0} -s^{1}\hat{p}^{1}+s^{2}\hat{p}^{2}+s^{3}\hat{p}^{3}
\right)\, ,
\\
\pm\tfrac{1}{4}
\left(
s_{0}\hat{q}_{0} +s^{1}\hat{p}^{1}-s^{2}\hat{p}^{2}+s^{3}\hat{p}^{3}
\right)\, ,
\\
\pm\tfrac{1}{4}
\left(
s_{0}\hat{q}_{0} +s^{1}\hat{p}^{1}+s^{2}\hat{p}^{2}-s^{3}\hat{p}^{3}
\right)\, ,
\\
\pm\tfrac{1}{4}
\left(
-s_{0}\hat{q}_{0} +s^{1}\hat{p}^{1}+s^{2}\hat{p}^{2}+s^{3}\hat{p}^{3}
\right)\, ,
\\
\pm\tfrac{1}{4}
\left(
-s_{0}\hat{q}_{0} -s^{1}\hat{p}^{1}+s^{2}\hat{p}^{2}+s^{3}\hat{p}^{3}
\right)\, ,
\\
\pm\tfrac{1}{4}
\left(
-s_{0}\hat{q}_{0} +s^{1}\hat{p}^{1}-s^{2}\hat{p}^{2}+s^{3}\hat{p}^{3}
\right)\, ,
\\
\pm\tfrac{1}{4}
\left(
-s_{0}\hat{q}_{0} +s^{1}\hat{p}^{1}+s^{2}\hat{p}^{2}-s^{3}\hat{p}^{3}
\right)\, ,
\\
\end{array}
\right.
\end{equation}

\noindent
for the $2^{4}$ possible choices of $s_{0},s^{1},s^{2},s^{3}$ of the charges in
Table~\ref{tab:stuextreme}. The first limit is $N=2$ supersymmetric etc. In all
cases, the extremal mass will be given by the same expression
Eq.~(\ref{eq:extremalmass}).

It is important to observe that the non-extremal solution has no constraints
on the signs (or the absolute values) of the charges, hence it interpolates
between the 16 discrete extremal limits.


\section{Conclusions}
\label{sec-conclusions}

In this paper we have constructed static non-extremal black-hole solutions of
three $N=2,d=4$ supergravity models using a general prescription based on
several well-known examples of non-extremal black holes. While we have given
some arguments to justify why this prescription may always work for all
models, we are far from having a general proof and more examples need to be
considered \cite{progress}.

On the other hand, the non-extremal solutions we have found are interesting per
se. They seem to share some important properties:

\begin{enumerate}

\item Even though in all the models considered there are several disconnected
  branches of extremal solutions, there is only one non-extremal solution that
  interpolates between all of them. All the extremal solutions are reachable
  by taking the appropriate extremal ($r_{0}\rightarrow 0$)
  limit. Furthermore, if we let $M$ diminish while leaving the charges and
  asymptotic values of the scalars constant (as happens in the
  evaporation process in these theories), which extremal limit is attained
  depends on the charges alone.

\item There seems to be a unique non-extremal superpotential in each theory
  and, in the different extremal limits, it gives the different superpotentials
  associated to the different branches of extremal solutions. 

\item The non-extremality parameter $r_{0}$, expressed in terms of the mass,
  charges and asymptotic values of the scalars, holds a great deal of
  information about the theory because $r_{0}$ vanishes whenever the value of
  the mass equals the value of any of the possible extremal superpotentials
  (some of which are the skew eigenvalues of the central charge
  matrix). Therefore, knowing this function
  $r_{0}(Z^{i}_{\infty},\mathcal{Q},M)$ we would know all the possible
  superpotentials. Unfortunately, there seems to be no a priori formula to
  determine it\footnote{Eq.~(\ref{eq:generalbound}) requires the knowledge of
    the scalar charges $\Sigma^{i}(Z^{i}_{\infty},\mathcal{Q},M)$, which we
    know how to compute only \textit{after} we have the complete black-hole
    solution.} and sometimes (e.g.~for the $STU$ model) it is not possible to
  find it explicitly even when the full solution is known.

\item The metrics have generically two horizons at the values $\tau=-\infty$
  (the outer, event horizon) and $\tau=+\infty$ (the inner, Cauchy horizon)
  whose areas and associated entropies are easily calculable and turn out to
  depend on the values of the scalars at infinity. The product of these two
  entropies is, in the three cases considered here, the square of the
  moduli-independent entropy of the extremal black hole that has the same
  charges.

\item The non-extremal solutions can be used to find some non-supersymmetric
  extremal solutions that cannot be constructed by the standard methods, as we
  have shown in the $\overline{\mathbb{CP}}^{n}$ model case.

\end{enumerate}

If this prescription works also in more complicated cases, it will give us the
opportunity to study how non-extremal black holes are affected by quantum
corrections and perhaps will give us new insights into the micrscopic
interpretation of the black-hole entropy in non-extremal cases. Work in this
direction is in progress.


\section*{Acknowledgments}

This work has been supported in part by the Spanish Ministry of Science and
Education grant FPA2009-07692, the Comunidad de Madrid grant HEPHACOS
S2009ESP-1473 and the Spanish Consolider-Ingenio 2010 program CPAN
CSD2007-00042. The work of CSS has been supported by a JAE-predoc grant
JAEPre 2010 00613. The work of PG has been supported in part by grants
FIS2008-06078-C03-02 and FPA2008-03811-E/INFN of Ministerio de Ciencia e
Innovaci\'on (Spain) and ACOMP/2010/213 from Generalitat Valenciana. TO wishes
to thank M.M.~Fern\'andez for her permanent support.


\end{document}